\documentclass[10pt,aps,prl,a4paper,twocolumn,longbibliography,superscriptaddress,floatfix]{revtex4-1}
\usepackage{bbold}
\usepackage{amsmath}
\usepackage{bm}
\usepackage{amssymb}
\usepackage[normalem]{ulem}
\usepackage{braket}
\usepackage{hyperref}
\usepackage{tikz}
\usepackage{todonotes}
\usepackage{ifthen}
\usetikzlibrary{tikzmark}

\usepackage{hyperref}
\hypersetup{%
   pdfpagemode=None, 
   pdfstartpage=1,
   pdfmenubar=true,
   pdftoolbar=true,
   colorlinks = true,
   linkcolor=blue,
   citecolor=blue,
   urlcolor=blue,
   bookmarksopen=false
 }

\def\ri{\mathrm i}
\def\re{\mathrm e}
\def\rd{\mathrm d}
\def\hc{{\rm h.c.}}
\def\tr{{\rm tr}}
\def\C{\mathcal{C}}
\def\DC{\Delta\C}
\def\GX{\Gamma X}
\def\DCav{\overline{\Delta\C}}
\def\up{\uparrow}
\def\down{\downarrow}

\def\MH{\mathcal{H}}
\def\omegaT{\tilde{\omega}}

\begin{document}

\title{Detecting topology through dynamics in interacting fermionic wires}

\author{Andreas Haller}
\affiliation{Institute of Physics, Johannes Gutenberg University, D-55099 Mainz, Germany}
\author{Pietro Massignan}
\email{pietro.massignan@upc.edu}
\affiliation{Departament de F\'isica, Universitat Polit\`ecnica de Catalunya, Campus Nord B4-B5, 08034 Barcelona, Spain}
\affiliation{ICFO – Institut de Ciencies Fotoniques, Barcelona Institute of Science and Technology, 08860 Castelldefels (Barcelona), Spain}
\author{Matteo Rizzi}
\affiliation{Forschungszentrum Jülich, Institute of Quantum Control, Peter Grünberg Institut (PGI-8), 52425 Jülich, Germany}
\affiliation{Institute for Theoretical Physics, University of Cologne, D-50937 K\"oln, Germany}

\begin{abstract}
Probing the topological invariants of interacting systems stands as a grand and open challenge. Here we describe a dynamical method to characterize 1D chiral models, based on the direct observation of time-evolving bulk excitations. We present analytical and state-of-the-art numerical calculations on various flavors of interacting Su-Schrieffer-Heeger (SSH) chains, demonstrating how measuring the mean chiral displacement allows to distinguish between topological insulator, trivial insulator and symmetry-broken phases. Finally, we provide a readily-feasible experimental blueprint for a model displaying these three phases and we describe how to detect those.
\end{abstract}

\maketitle

Since the discovery of sharp resistivity jumps in Quantum Hall devices and their explanation in terms of quantized invariants, topology played an ever increasing role in modern condensed matter \cite{Hasan2010,Bernevig2013}. Transport experiments are often employed in solid state systems to observe quantized responses, but in ultracold atoms, polaritonic and photonic systems other avenues are generally pursued, such as direct imaging of edge states, detection of anomalous displacement or interferometry \cite{Cooper2019,Ozawa2019}.
Moreover, detecting interaction-induced topological transitions in such setups stands as an open and urgent issue.

Topological invariants are most simply defined as integrals over momentum space.
In one-dimensional chiral models, for example, the relevant invariant is the winding number of a vector in the two-dimensional space dictated by the symmetry~\cite{Asboth2016}.
In presence of interactions this winding number can be defined as the chiral winding of the zero-frequency component of the Fourier-transformed single-particle imaginary-time Green's function ${g\equiv G(k,\omega=0)}$, via~\cite{Gurarie2011,Manmana2012}
\begin{align}\label{eq:winding_gurarie}
    \gamma = {\rm tr}\int \frac{\rd k}{4\pi\ri}\,\Gamma g^{-1} \partial_k g\,,
\end{align}
with $g^{-1}$  the matrix inverse of $g$, and $\Gamma$ the chiral symmetry operator, which anti-commutes with the Bloch Hamiltonian if this symmetry is present~\footnote{A chiral symmetry in condensed matter is commonly defined in terms of an {\it anti-unitary} operator {$\widehat{\Gamma}$}, acting {\it locally} on the fermionic operators $c_{x,\tau,s}$ as {$\widehat{\Gamma} c_{x,\tau,s} \widehat{\Gamma}^{-1} = \Gamma_{\tau s,\tau' s'}c_{x,\tau',s'}^\dag$} with $\Gamma$ an even-dimensional {\it unitary} matrix squaring to the identity. 
The commutation of {$\widehat{\Gamma}$} with the Hamiltonian $\MH$ is equivalent to the anti-commutation of $\Gamma$ with the Bloch Hamiltonian $H(k)$.}.
For non-interacting systems with Hamiltonian $H_0(k)$ the Green's function is $G(k,\omega)=[\ri \omega-H_0]^{-1}$, so that $g=H_0^{-1}$ and Eq.~\eqref{eq:winding_gurarie} reduces to the usual winding of the Bloch Hamiltonian.

Accessing momentum space to compute such invariants in experiments is possible \cite{Atala2013}, but still not always evident.
Obvious examples are systems with broken translational invariance.
Alternative route discussed in the literature include flux insertions \cite{Niu1985,Xiao2010,Altland2014,Altland2015} and field theory anomalies \cite{Ryu2012}.
In real space, spectral projectors have been exploited for studying topological invariants in Chern insulators~\cite{Bianco2011}, disordered systems \cite{Mondragon-Shem2014}, quasicrystals \cite{Tran2015}, and non-Hermitian models~\cite{Rudner2009,Zeuner2015}.

Another method which proved very successful to detect the topology of single-particle 1D chiral models is the read-out of the {\it mean chiral displacement} (MCD), which can be defined in a many-body context as
\begin{align}\label{eq:mcd_definition}
    \mathcal C(t) =
        \sum_{\tau,s}
        \Big\langle
            c^{}_{0,\tau,s} \re^{\ri\MH t/\hbar} \GX \re^{-\ri\MH t/\hbar} c^\dag_{0,\tau,s}
        \Big\rangle
        \,.
\end{align}
In the above, $c^\dagger_{x,\tau,s}$ is the creation operator of a fermionic particle (with sublattice index $\tau$ and spin $s$ acting on the unit-cell at position~$x$), lattice spacing is set to unity, and $\Gamma X = \sum_{x,\tau,\tau',s,s'} c^\dag_{x,\tau,s} (x \Gamma^{}_{\tau s,\tau' s'}) c^{}_{x,\tau',s'}$.
The mean chiral displacement thus measures the propagation of an excitation created at the central unit-cell $x=0$ after an evolution time $t$, weigthed by the chiral operator $\Gamma$.
In Refs.~\cite{Cardano2017,Meier2018,Maffei2018} it was shown theoretically and confirmed experimentally that the MCD of a localized single-particle excitation above the vacuum state converges in the long-time limit to the chiral invariant of one dimensional static, periodically-driven, and even disordered systems.
Since then, the MCD has found a myriad of applications in very different single-particle systems \cite{Zhou2018,Wang2018,Xie2019,Bomantara2019,Wang2019,Xie2019b,Zhou2019,DErrico2020,StJean2020}.

In this work, we prove that the mean chiral displacement is an ideal candidate to detect topological features of many-body interacting models in current experiments.
First, we show that the single-particle (disconnected) part of the MCD, measured after sudden creation of local excitations, coincides precisely with the many-body winding defined in Eq.~\eqref{eq:winding_gurarie}.
This implies that measuring the MCD on top of a half-filled Fermi sea provides a direct and exact detection of the winding.
Then, we present nonperturbative matrix product states (MPS) simulations of a versatile 1D chiral model.
In a weakly-correlated configuration, the connected part of the MCD remains negligible, and the MCD keeps converging reliably to the winding number.
Switching instead to strong many-body correlations, we show that the MCD also signals emergent symmetry-breaking long-range ordered phases.
We conclude by discussing a blueprint scheme for the experimental detection of the MCD on a strongly-correlated model.

\begin{figure}[ht]
    \includegraphics[trim={0cm 4.2cm 0 0},clip]{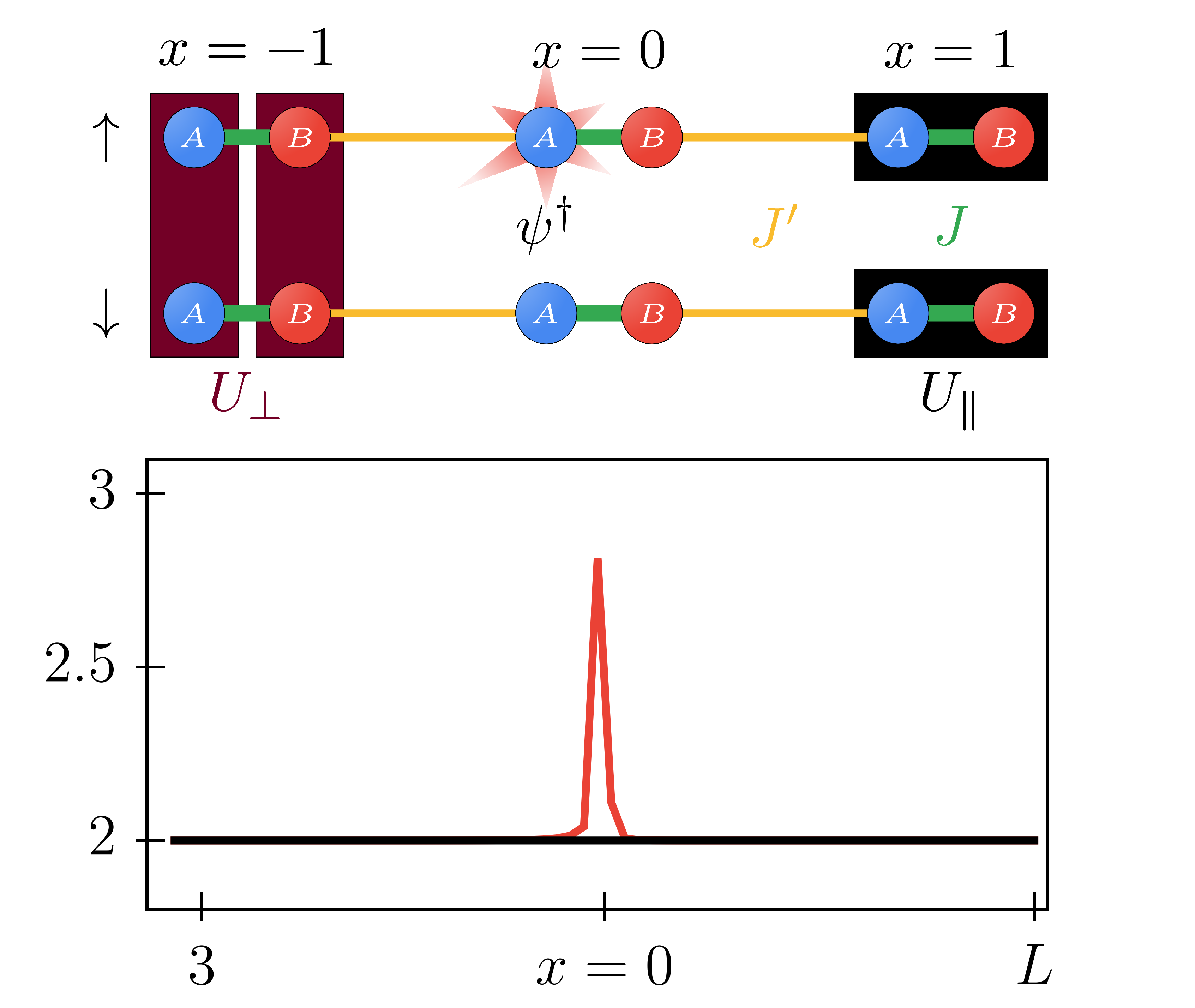}\llap{\parbox[b]{15.75cm}{(a)\\\rule{0ex}{2.5cm}}}
    \\
    \includegraphics{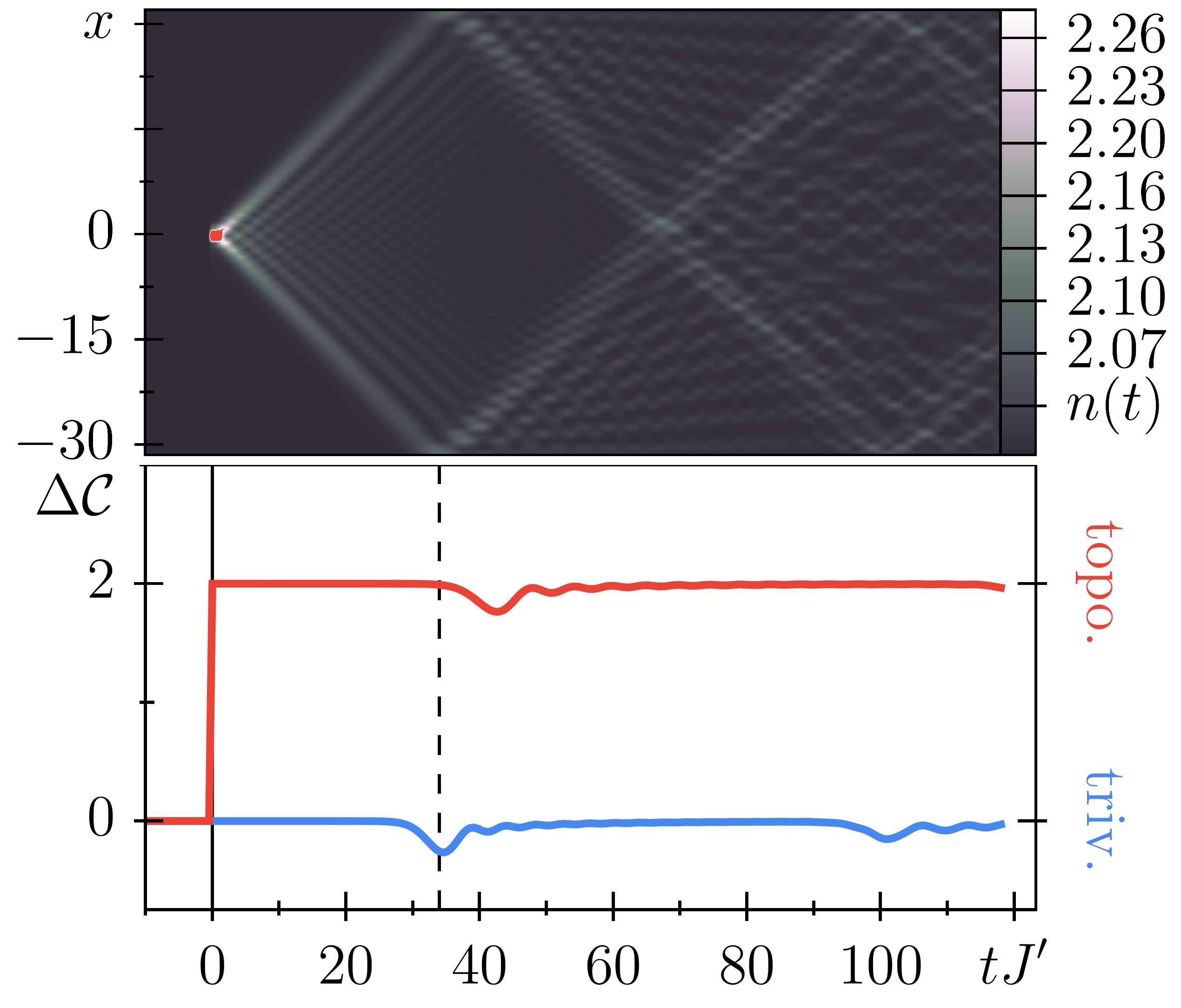}\llap{\parbox[b]{15.75cm}{(b)\\\rule{0ex}{6.5cm}}}\llap{\parbox[b]{15.75cm}{(c)\\\rule{0ex}{2.1cm}}}
    \caption{
        (a) Sketch showing three unit-cells of a spinful SSH chain with sublattices $A$ and $B$, spin states $\up$ and $\down$, intra- and inter-cell tunnelings $J$ and $J'$, and interactions $U_\parallel$ and $U_\perp$.
        (b) Dynamics of a local excitation: adding one extra fermion at ($t=0,\,x=0$) to a half-filled single-component Fermi sea generates an evolution with a characteristic cone-like spreading pattern (model: non-interacting SSH chain with $L=64$ unit-cells, $J/J'=1.2$ and open boundary conditions).
        (c) Mean chiral displacement for $J/J'=0.8$ (red, $\gamma=2$) and $1.2$ (blue, $\gamma=0$). In non-interacting systems, the MCD is quantized to the winding number. Minute oscillations appear (and relax quickly) upon scattering with the chain boundaries, regardless of the presence or absence of edge states.
    }
    \label{fig:recipe}
\end{figure}

\begin{figure}[t]
    \includegraphics[width=0.493\columnwidth]{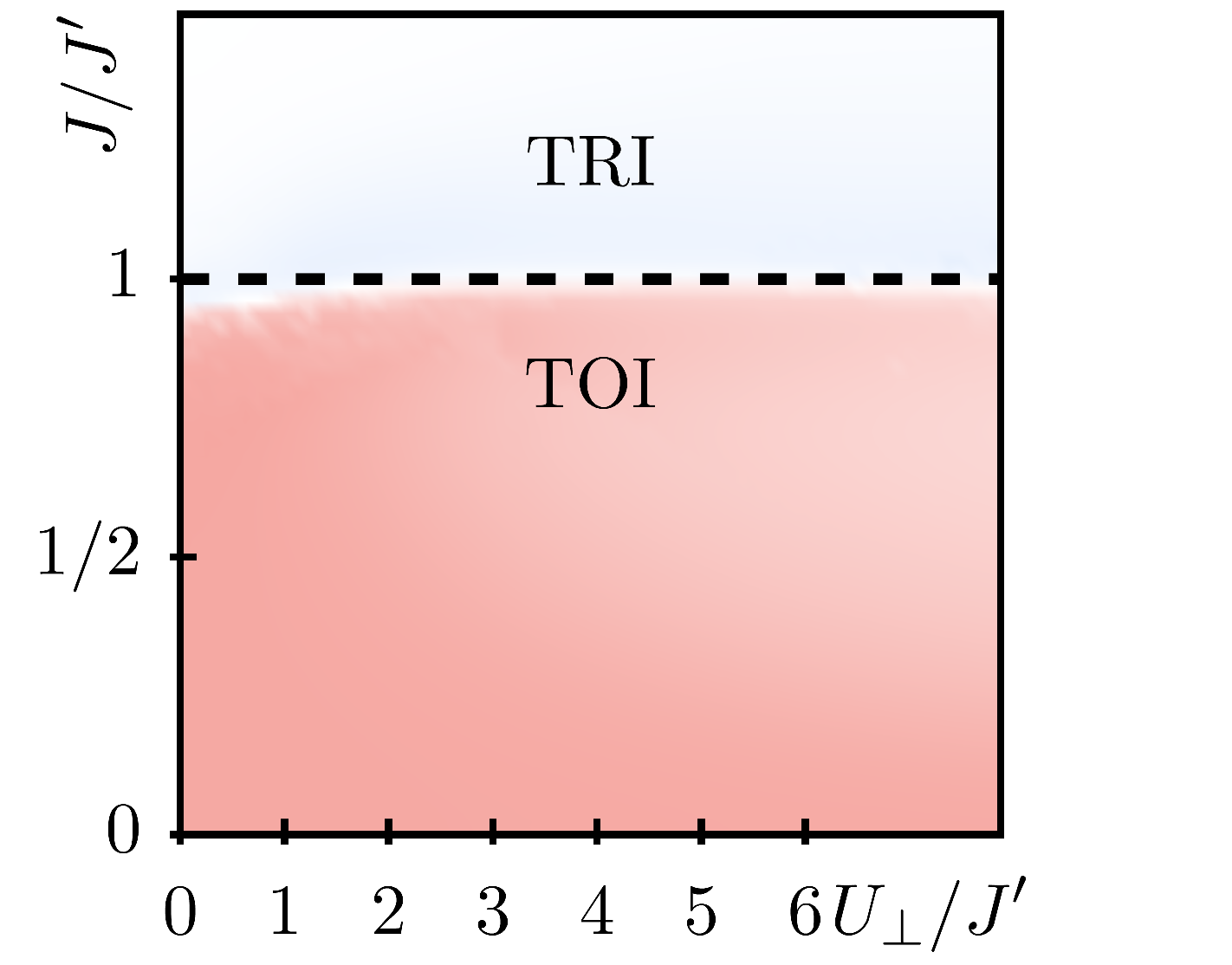}\llap{\parbox[b]{2.5cm}{(a)\\\rule{0ex}{2.6cm}}}
    \includegraphics[width=0.493\columnwidth]{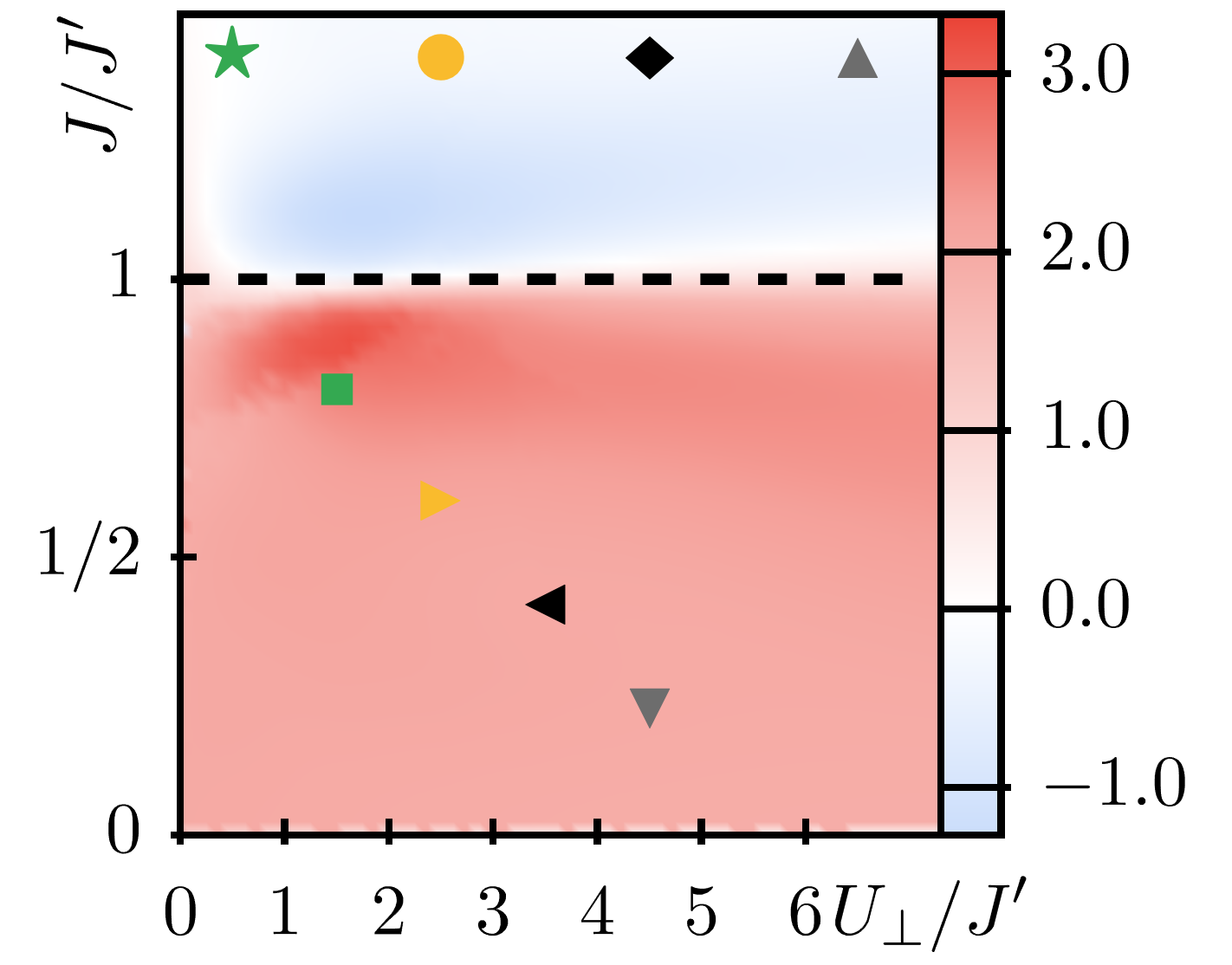}\llap{\parbox[b]{2.5cm}{(b)\\\rule{0ex}{2.6cm}}}
    \includegraphics[width=0.493\columnwidth]{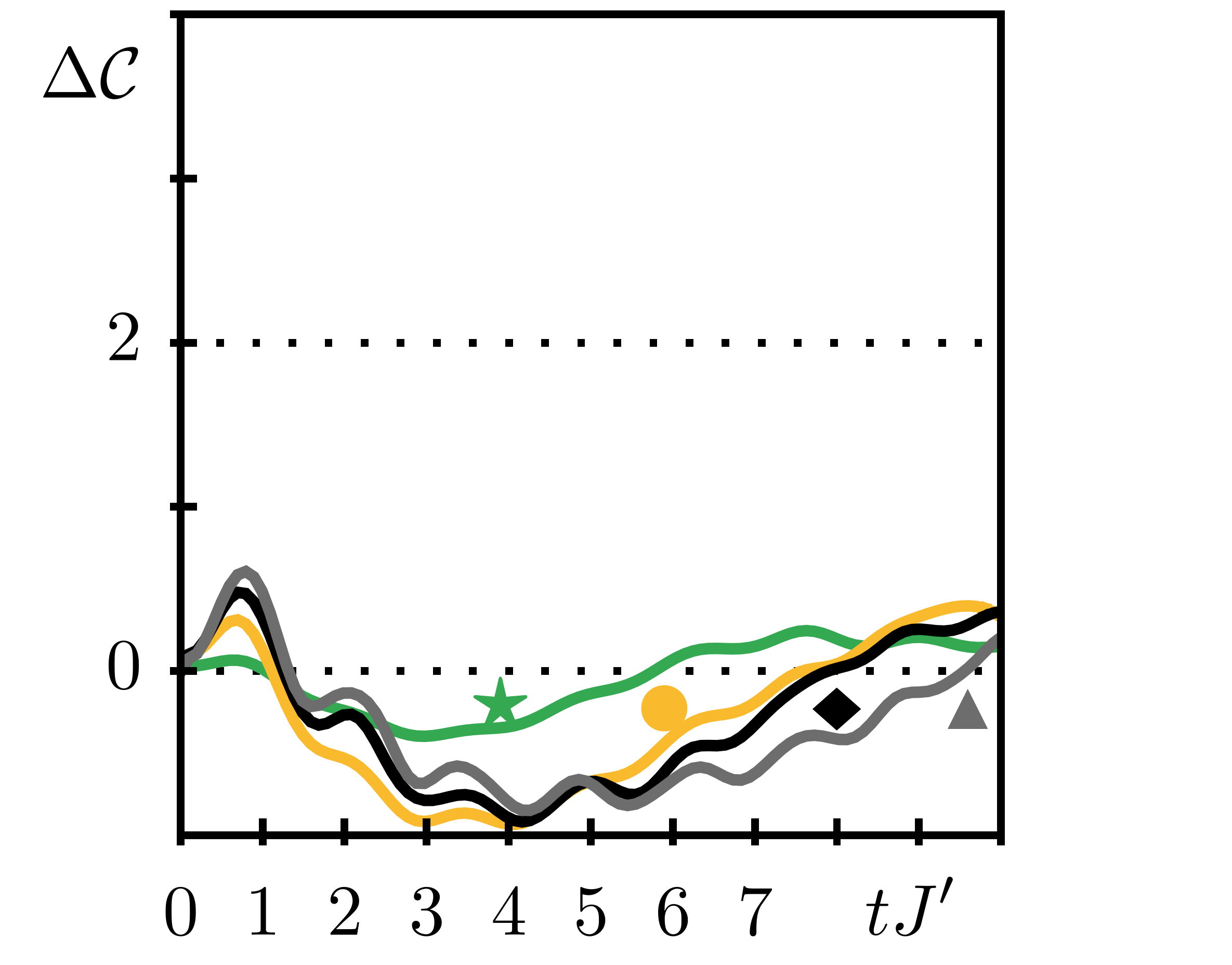}\llap{\parbox[b]{2.6cm}{(c)\\\rule{0ex}{2.7cm}}}
    \includegraphics[width=0.493\columnwidth]{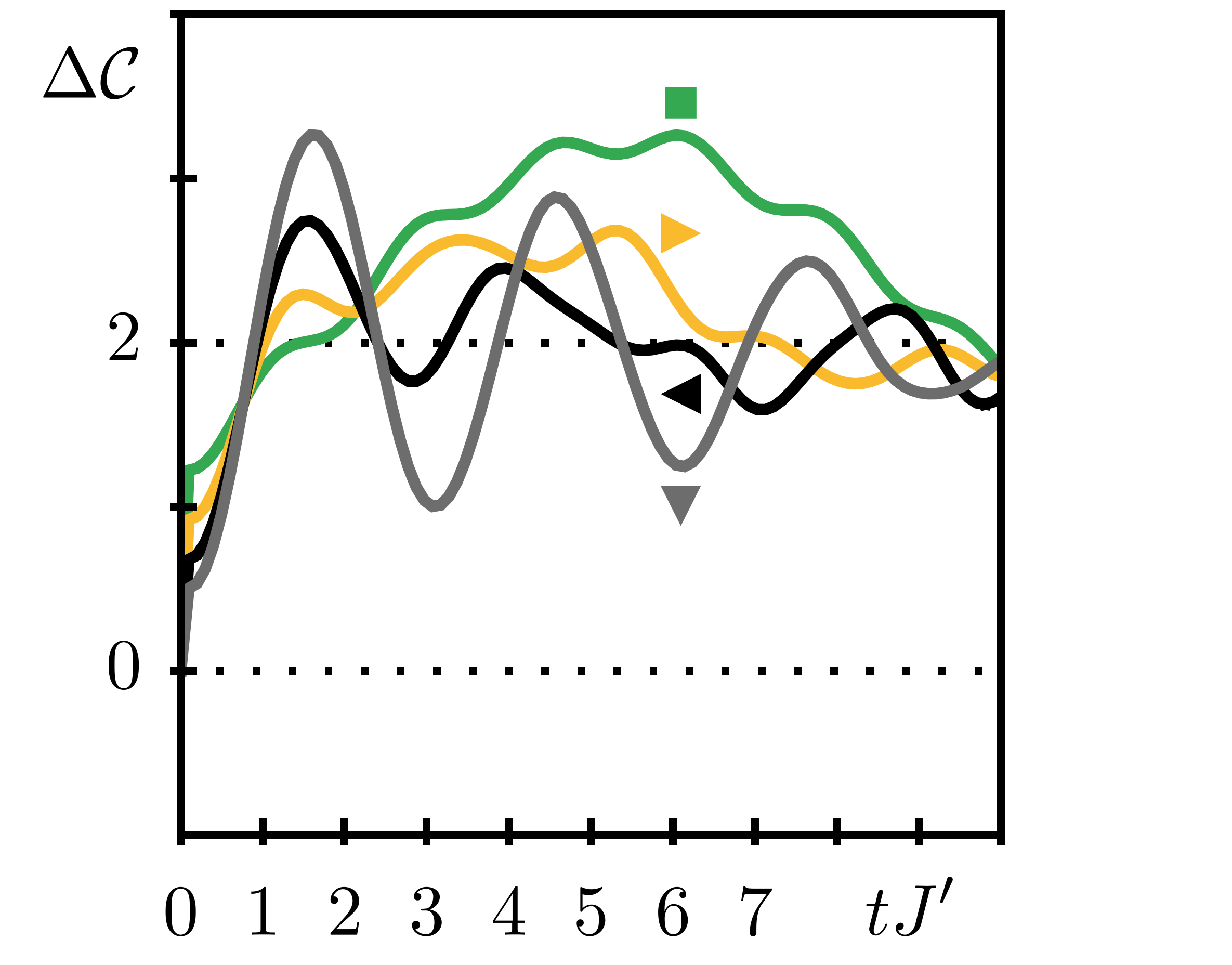}\llap{\parbox[b]{2.6cm}{(d)\\\rule{0ex}{2.7cm}}}
    \caption{MPS analysis of the MCD of the short-ranged Hamiltonian $\MH_{\rm sr}$ in a system with $L=32$ unit-cells and open boundary conditions.
    The phase diagram contains trivial (TRI) and topological insulator (TOI) phases, separated by a transition (dashed line) independent of $U_\perp$.
    (a) Time-averaged disconnected part $\DCav_{\rm d}$ of the MCD.
    (b) Time-averaged full correlator $\DCav$.
    (c-d) Detailed (i.e., un-averaged) time-traces of $\Delta \mathcal C(t)$ at parameters marked by colored symbols in panel (b)  are shown in panels (c) and (d).
    Both here and in Fig.~\ref{fig:mcd_numerics_lr}, the time averages in panels (a) and (b) are performed over the time-duration $T=10/J'\ll L/J'$, which ensures that the excitation propagates through the bulk only, without reaching the edges of the chain.
    }
    \label{fig:mcd_numerics_sr}
\end{figure}

\begin{figure}[ht]
    \includegraphics[width=0.493\columnwidth]{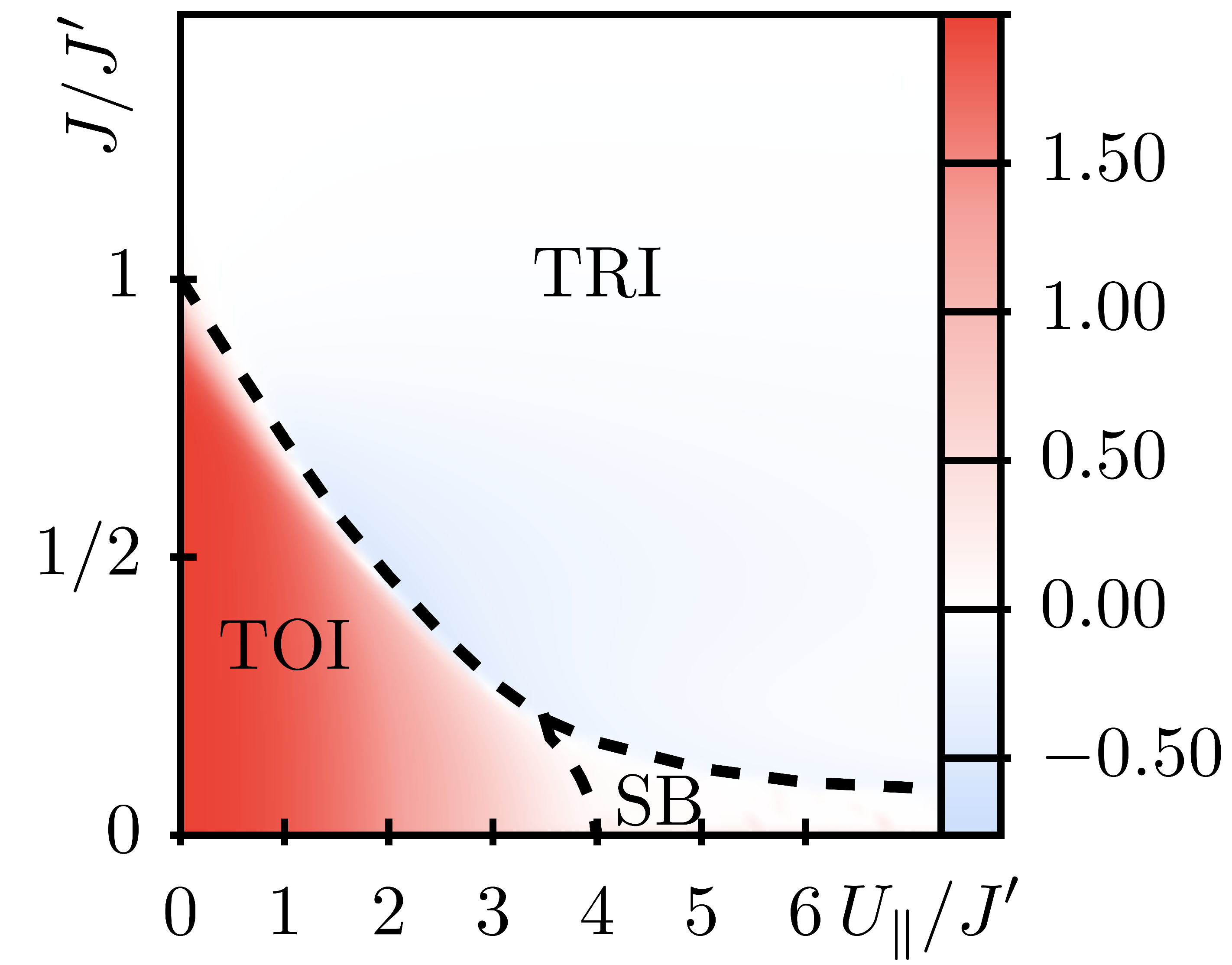}\llap{\parbox[b]{2.5cm}{(a)\\\rule{0ex}{2.7cm}}}
    \includegraphics[width=0.493\columnwidth]{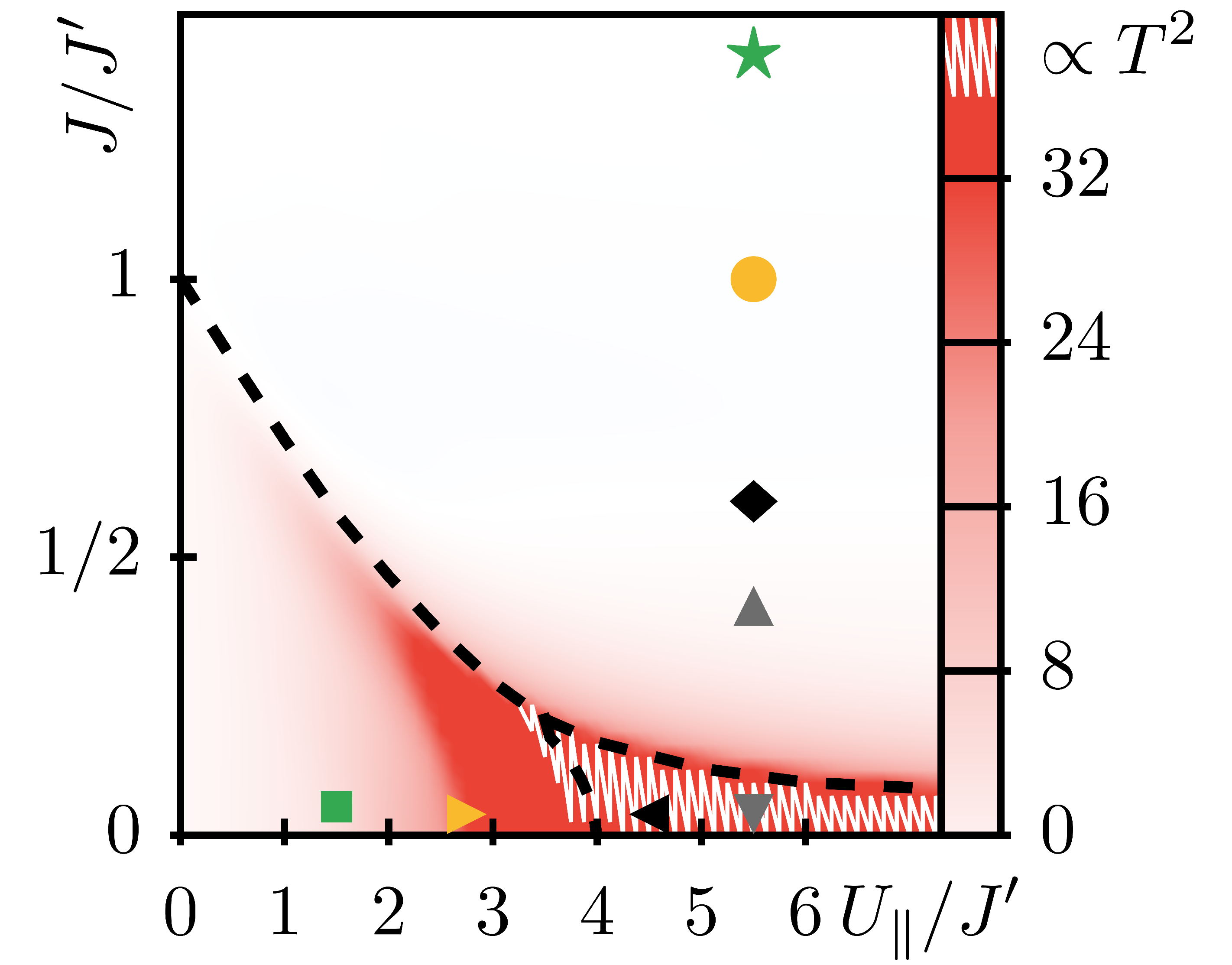}\llap{\parbox[b]{2.5cm}{(b)\\\rule{0ex}{2.7cm}}}
    \includegraphics[width=0.493\columnwidth]{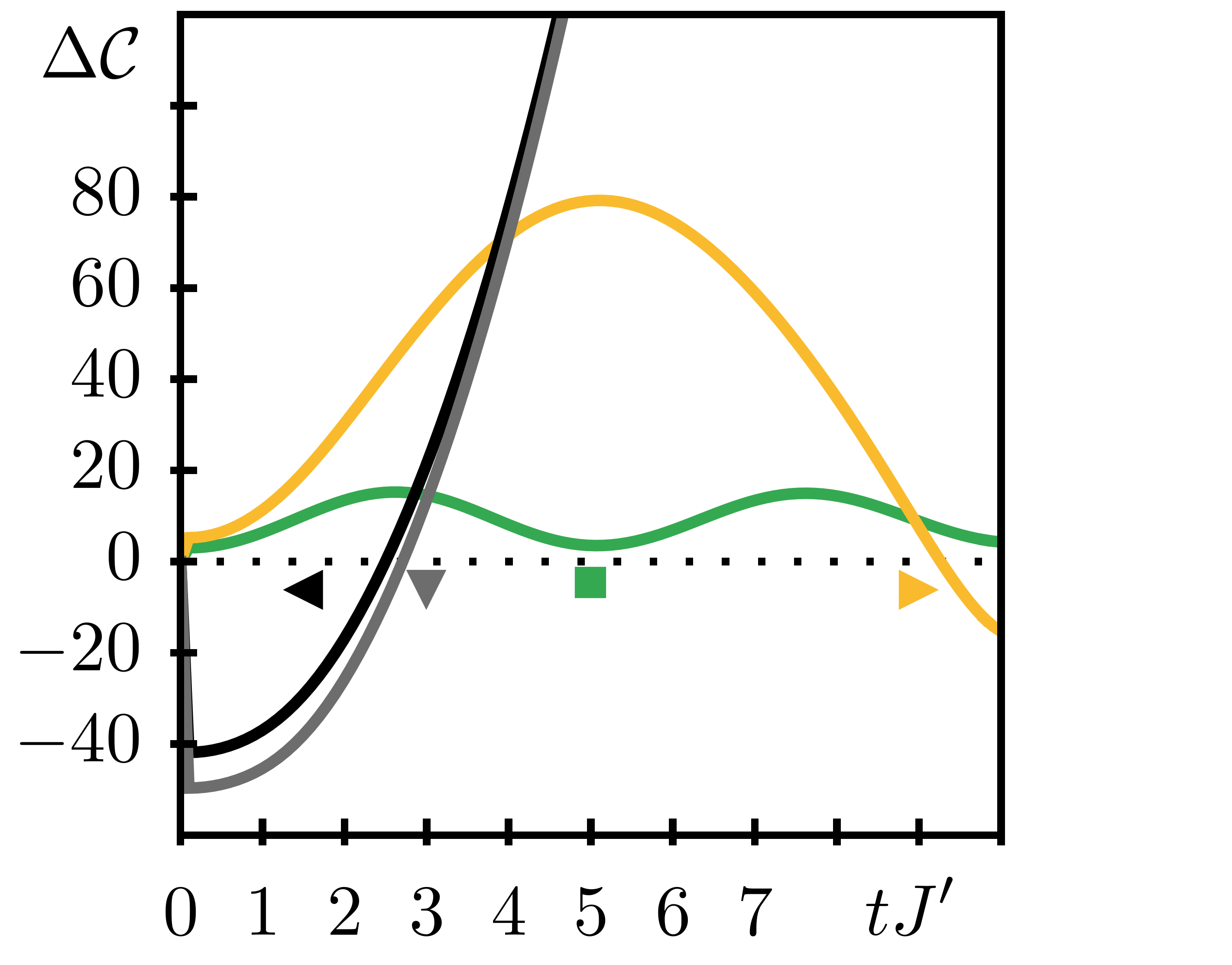}\llap{\parbox[b]{2.5cm}{(c)\\\rule{0ex}{2.7cm}}}
    \includegraphics[width=0.493\columnwidth]{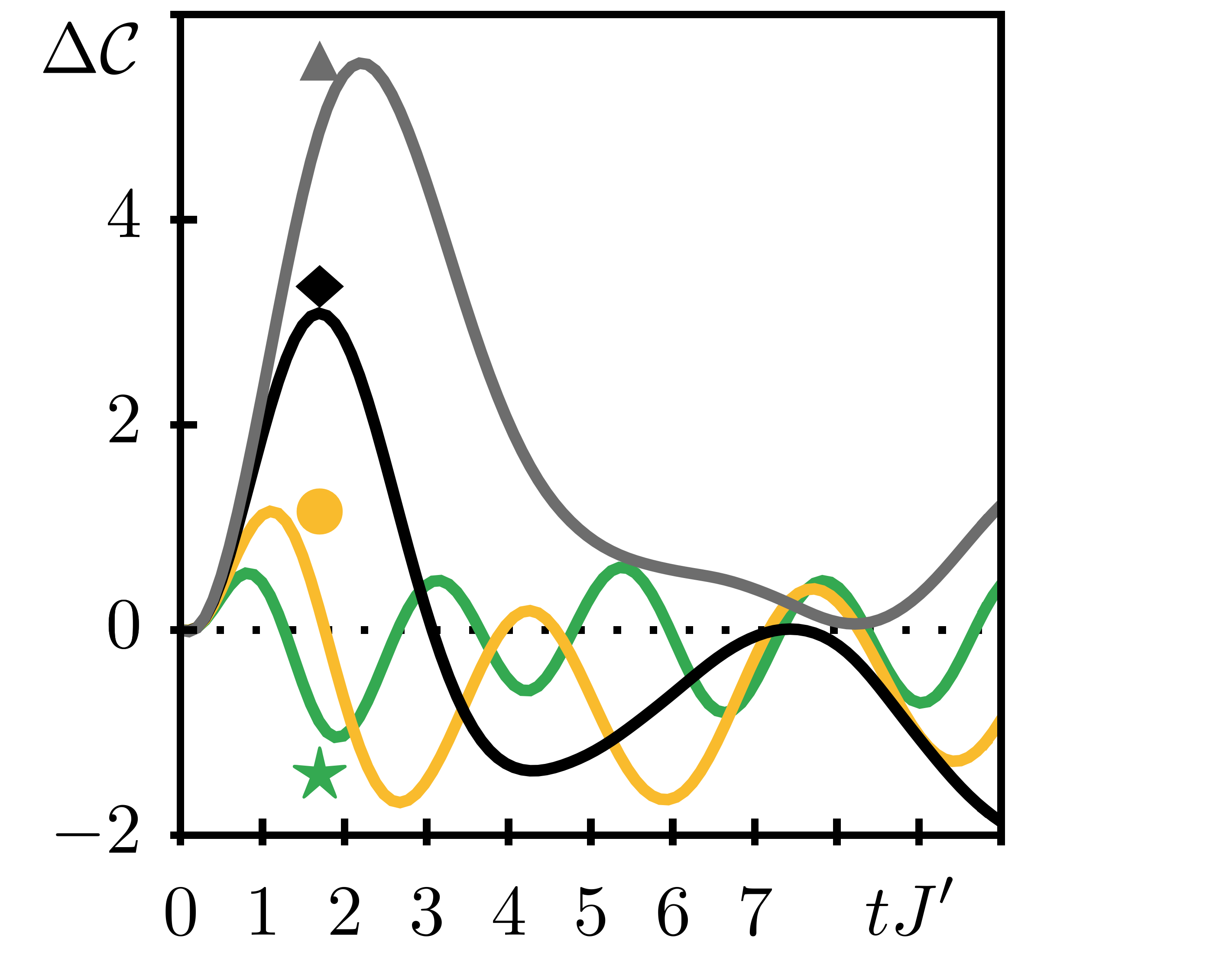}\llap{\parbox[b]{2.5cm}{(d)\\\rule{0ex}{2.7cm}}}
    \caption{MPS analysis of the long-ranged Hamiltonian $\MH_{\rm lr}$ for $L=32$ unit-cells and open boundary conditions. Aside from trivial (TRI) and topological insulator (TOI) regions, the phase diagram includes an extra symmetry-breaking phase (SB).
    (a) The disconnected part $\DCav_{\rm d}(T)$ of the MCD.
    (b) Time-average $\DCav(T)$ of the full correlator. The hatched region with zigzag lines denotes $\DCav\propto T^2$. Details of the dynamics of $\Delta \mathcal C$ at parameters marked by colored symbols in panel (b)  are shown in panels (c) and (d).
    Although the time-averaged MCD for a finite system does not distinguish well between SB and TOI, the behavior of $\DC$ is markedly different in all three regions: it oscillates around the initial value in the TRI, around a finite (but not quantized) value in the TOI, and it diverges quadratically in the SB phase.
    Close to phase transitions, a finite-size scaling lifts the ambiguity between the different cases~(see Fig.~\ref{fig:MPS_long_time}).
    }
    \label{fig:mcd_numerics_lr}
\end{figure}

{\em Mean chiral displacement as a topological marker.--}
Let us start by clarifying the relation between the many-body topological invariant $\gamma$ defined via single-particle Green's functions and the MCD.
For this purpose, let us consider a system in its ground state for times $t< 0$, and excite it by adding one particle at $(x=0,t=0)$.
The time dependence of the MCD evaluated on the perturbed ground state at half filling can be expanded (for details see Supplemental Material~\cite{SM}) as
\begin{align}\label{eq:MCD_wicks_theorem}
    \mathcal C(t) = \xi(t) + \sum_{\tau,s}\langle1-n^{}_{0,\tau,s}\rangle\langle\GX\rangle + \sum_x x\,\tr\left(G^\dag\Gamma G\right)
\end{align}
with local density $n_{x,\tau,s}=c^\dag_{x,\tau,s}c^{}_{x,\tau,s}$, averages $\langle\ldots \rangle$ evaluated on the unperturbed ground state, $\xi(t)$ the connected (many-body) part of $\C(t)$ and $G$ the one-body Green's function  matrix, with components ${\ri G_{\tau s,\tau' s'}(x,t) = \langle c^{}_{x,\tau,s\vphantom'}(t)c^\dag_{0,\tau',s'}(0)\rangle}$ ($t>0$).
The second sum is constant over time, so that the time-dependence of the MCD is given by
\begin{align}\label{eq:winding_MCD}
    \DC(t) \equiv \C(t) - \C(0^-) = \sum_x x\cdot\tr\left(G^\dag\Gamma G\right) + \Delta\xi(t)
\end{align}
with $\Delta\xi(t)=\xi(t) - \xi(0^-)$ the many-body contribution to the time-dependent MCD, where $0^-$ indicates the time before creating the excitation.
For later convenience, we define the one-body contribution $\DC_d(t) = \DC - \Delta\xi$.
In a non-interacting many-body system (e.g., a Fermi sea), the many-body part $\Delta\xi$ of the correlator is zero and the only remainder is the one-body part.
When a single particle is added to a half-filled Fermi sea, the MCD is remarkably {\em quantized}, and equal to the topological invariant (see Fig.~\ref{fig:recipe}(c), and a detailed proof in~\cite{SM}).
When instead excitations are generated within an interacting system, or within a Fermi sea away from half-filling, scattering processes between conduction and valence band give rise to damped oscillations, or more complex behaviors. To filter away oscillatory contributions, it is useful to define the time-averaged MCD 
$\DCav \equiv\frac{1}{T}\int_0^T{\rm d}t\, \DC(t)$.
In the following we show by nonperturbative MPS simulations~\cite{Haegeman2011,Paeckel2019,Calabrese2005,Calabrese2007} presented in Figs.~\ref{fig:mcd_numerics_sr} and~\ref{fig:mcd_numerics_lr}, that the time-average of the one-body (disconnected) term
\begin{align}\label{eq:winding_MCD_2}
    \DCav_d = \frac{1}{T}\int_0^T{\rm d}t\,\sum_x x\cdot\tr\left(G^\dag\Gamma G\right)
\end{align}
provides an approximation of the winding number $\gamma$ defined in real space and real time.
Moreover, we show that the MCD is also a useful marker of spontaneous symmetry-breaking in trivial insulators, as the latter gives rise to a divergent many-body term $\xi(t)$.

{\em Model: Interacting SSH chains.--}
To illustrate our findings we focus on a specific 1D model, the interacting Su-Schrieffer-Heeger (SSH) fermionic chain, which is shown schematically in Fig.~\ref{fig:recipe}(a). The Hamiltonian of the model reads $\MH = \MH_0 + \MH_\perp + \MH_\parallel$.
The kinetic part $\MH_0$ is a tight binding model with two sublattices ($\tau\in\{A,B\}$) and internal spin-1/2 ($s\in\{\up,\down\}$) degree of freedom denoted by
\begin{align}\label{eq:tight_binding}
    \MH_0 = J \sum_{x,s} c^\dag_{x,A,s} c^{}_{x,B,s} + J'\sum_{x,s} c^\dag_{x,B,s} c^{}_{x+1,A,s} + \hc  
\end{align}
The intra-cell and inter-cell tunneling amplitudes are denoted by $J$ and $J'$, respectively.
In the following, we will set $\hbar=1$, and time will be measured in units of $1/|J'|$.
On top of the kinetic terms we consider Hubbard on-site interactions between $\up$ and $\down$ spins,
\begin{align}\label{eq:H_U}
    \MH_\perp  =U_\perp \sum_{x,\tau}\left(n_{x,\tau,\up}-\frac12\right)\left(n_{x,\tau,\down}-\frac12\right)\,,
\end{align}
and interactions which act between two identical spins located on the $A$ and $B$ sites of the same unit-cell,
\begin{align}\label{eq:H_para}
    \MH_\parallel = U_\parallel\sum_{x,s}\left(n_{x,A,s}-\frac12\right)\left(n_{x,B,s}-\frac12\right)\,,
\end{align}
Note that $\MH_0$ is fully decoupled in the spin-sector, and, when considering only $\MH_0 + \MH_\parallel$, it is redundant to account for both spin orbitals.

{\em The non-interacting case.--}
Let us first review the topological properties of $\MH_0$.
This 1D Hamiltonian has chiral symmetry,  which places it in the AIII class of topological insulators \cite{Chiu2016}.
If moreover $J$ and $J'$ are real, $H_0$ features particle-hole and time-reversal symmetries,
and the model belongs to the more restrictive class BDI.
For each of the two disconnected SSH chains, the Bloch Hamiltonian winds once or not at all around the origin, so that $\gamma=2$ for $|J/J'|<1$ and $\gamma=0$ for $|J/J'|>1$.
This is particularly easy to see in two extreme limits.
When $J'=0$, unit-cells are disconnected and $G_{As,Bs'}(x,t)$
is zero unless $x=0$, thus $g_{As,Bs'}(k)$ is momentum-independent and with zero winding.
In the other extreme case $J=0$, we have $g_{As,Bs'} = g_0\delta_{s,s'} \re^{\ri k}$, which (summing over the spin degrees of freedom) yields a winding of $2$~\cite{Manmana2012}.
In Fig.~\ref{fig:recipe}(b) we evaluate the dynamics of excitations on top of a half-filled Fermi sea by exact diagonalization of the quadratic Hamiltonian (see~\cite{SM} for details), and we show that the corresponding MCD captures correctly the winding.

{\em The short-ranged case.--}
Let us now consider the Hamiltonian ${\mathcal{H}_{\rm sr}=\MH_0 + \MH_\perp}$, also known as {\it Peierls-Hubbard model}, whose topological properties were first studied in Ref.~\cite{Manmana2012}.
When $|J/J'|<1$ the system remains topological for {\em every} value of $U_\perp$ because the interaction terms in $\MH_\perp$ do not couple edge with bulk modes and therefore zero-energy edge excitations remain unaffected \cite{Manmana2012}.
At $|J|=|J'|$, the system is a 1D Hubbard model: while a charge gap is immediately opened by $U_\perp>0$, the spin sector remains gapless and thus provides the phase boundary between topological and trivial insulator.
Its dimerized- and bond-limits ($J'=0$ and $J=0$, respectively) are short-range correlated and thus we expect that $\DC\approx\DC_{\rm d}$ will oscillate around the winding number $\gamma$.
In Fig.~\ref{fig:mcd_numerics_sr} we confirm this by computing the MCD and its disconnected part with MPS simulations.
Our findings agree with the results of Ref.~\cite{Manmana2012}, and here we also provide the missing link to an observable which is easily accessible in ongoing experiments based on fermionic quantum gas microscopes \cite{Haller2015,Cheuk2015,Parsons2015,Omran2015} or optical tweezers \cite{Murmann2015}.

{\em Effective interacting spin-model.--}
To proceed further, we introduce an alternative representation of the Hamiltonian.
This will allow us to illustrate the crucial differences between $\MH_\parallel$ and $\MH_\perp$ interactions on the topological properties of the underlying system.
Let us rewrite the model in the {\it link-basis} defined by $\ell_{x,+,s} = (c_{x+1,A,s} + c_{x,B,s})/\sqrt2$ and $\ell_{x,-,s}=(c_{x+1,A,s} - c_{x,B,s})/\sqrt2$.
By introducing the spinor $D_{x,s}=(\ell_{x,+,s},\ \ell_{x,-,s})^T$ we can define a set of pseudospin operators $S^i_{x,s} = D_{x,s}^\dag \sigma_i D_{x,s}^{\vphantom\dag}/2$ in which $\sigma_i$ denotes the Pauli matrix $i\in\{x,y,z\}$.
These are particularly useful in the limit $J=0$, because the individual terms of $\MH$ become expressions in $S^i_{x,s}$
that all clearly commute with each link-density $n_{x+1,A,s}+n_{x,B,s}$ separately.
For a translationally invariant system at half-filling, this implies
$n_{x+1,A,s}+n_{x,B,s}=1\,\, \forall x,s$, with effective Hamiltonian
\begin{align}\label{eq:hamiltonian_pseudospins}
    \MH_{\rm eff}\big|_{J=0} &=
    2J'\sum_{x',s} S^z_{x',s}
    +2U_\perp\sum_{x'}\left(\frac14 + S^x_{x',\up}S^x_{x',\down}\right)
    \nonumber\\
    &\qquad+U_\parallel\sum_{x',s}\left(\frac14-S^x_{x'-1,s}S^x_{x',s}\right).
\end{align}
As a consequence, ${\MH_0 + \MH_\perp}$ results in a product state with anti-aligned pseudospins at each unit-cell, which implies a many-body state with short-range correlations.
Strong correlations are instead induced by the interplay between $\MH_0$ and $\MH_\parallel$, resulting in two copies of a transverse-field Ising model with quasi long-range ferromagnetic ordering if $U_\parallel>4|J'|$~\cite{Juenemann2017,Tirrito2019}.
A more detailed derivation of this model may be found in the Supplemental Material~\cite{SM}.

{\em The long-ranged case.--}
Let us now consider $\mathcal{H}_{\rm lr}=\MH_0 + \MH_\parallel$.
The two spin states are effectively decoupled, so that we focus only on a single chain, for which the winding $\gamma$ may be either 0 or 1.
As shown in the Supplemental material, an appropriate rotation maps $\mathcal{H}_{\rm lr}$ to the Creutz model considered in Ref.~\cite{Creutz1999,Juenemann2017,Tirrito2019}.
Its phase diagram features an interaction-induced transition from topological (TOI) to trivial insulator (TRI), which is readily detectable by the disconnected part of the MCD -- we present MPS simulations thereof in Fig.~\ref{fig:mcd_numerics_lr}(a). Correspondingly, the time-averaged MCD shown in Fig.~\ref{fig:mcd_numerics_lr}(b) goes smoothly to zero in the trivial region, while it saturates to a finite value in the topological region.
Interestingly, at strong $U_\parallel$  interactions an exotic interaction-induced phase transition can also emerge.
The ferromagnetic phase arises from a spontaneous symmetry breaking (SB) of the Ising $\mathbb Z_2$ symmetry, and corresponds to the Aoki phase of the Gross-Neveu model on a lattice~\cite{Bermudez2018,Tirrito2019,Kuno2019}.
Indeed the excitation generates the appearance of an anti-aligned domain in the local $A,B$ density, whose size grows linearly in time.
In turn, this implies that the corresponding MCD grows quadratically in time -- as it is clearly visible in our simulations in the ``SB" region [see gray and black traces in Fig.~\ref{fig:mcd_numerics_lr}(c)-(d)].

\begin{figure}[ht]
    \includegraphics{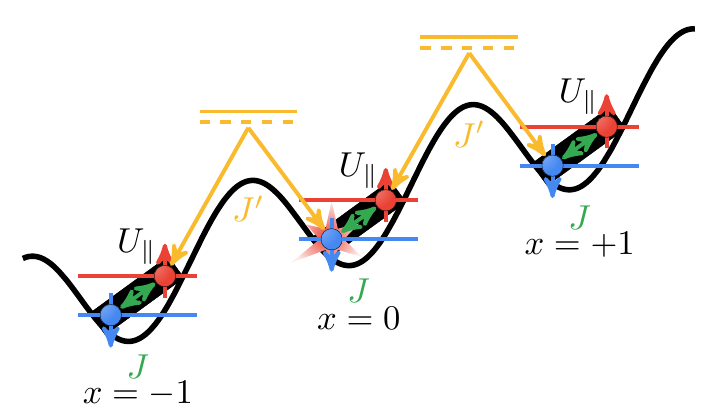}
    \caption{Proposal to engineer the long-range Hamiltonian $\MH_{\rm lr}=\MH_0 + \MH_\parallel$ with two-component ultracold fermions in a tilted optical lattice. A Zeeman term splits the degeneracy between the $\up$ and $\down$ states. A linear tilt is used to hinder particles from naturally hopping between lattice sites.
    The desired hopping processes are reintroduced by i) RF transitions between the $\uparrow/\downarrow$ states (green), and ii) two-photon Raman transitions (yellow).
    To create a hole excitation, a blast beam tightly-focused on $x=0$ transfers one particle to an internal state which is uncoupled to the Hamiltonian.}
    \label{fig:experiment}
\end{figure}

{\em Experimental blueprint for $\MH_{\rm lr}$.--}
The intriguing phase diagram shown in Fig.~\ref{fig:mcd_numerics_lr} is readily accessible through experiments with ultracold atoms trapped in optical lattices.
Indeed, in Fig.~\ref{fig:experiment} we present a simple proposal yielding the long-ranged Hamiltonian $\MH_{\rm lr}=\MH_0 + \MH_\parallel$, based on two-component fermions in a tilted lattice, exposed to RF and two-photon Raman transitions.
In this implementation, spin $(\down/\up)$ and sublattice $(A/B)$ degrees of freedom are effectively identified.
To perform the measurement, the system should be prepared in the model's ground state, i.e., at half-filling.
Subsequently a localized excitation $|e\rangle$ is created at position $x=0$ and arbitrary sublattice.
The easiest strategy here would be to create a quasi-hole, by ``blasting" away one particle by means of a focused excitation.
Successive readouts of the spin-resolved average positions $X_\down=\sum_x x {\langle n_{x,\down}(t)\rangle_e}$ and $X_\up=\sum_x x \langle n_{x,\up}(t)\rangle_e$ (where ${\langle \ldots\rangle_e}$ denotes an average over the excited state created by $|e\rangle$)
are used to compute the MCD, which in this case very simply coincides with the difference $\C(t)=\pm (X_\down - X_\up)$.
Here, $\pm$ depends on whether $|e\rangle$ adds or removes a particle~\cite{SM}.
For a detection of topological phase diagrams in general, a read-out of the disconnected part $\DC_d$ is possible by means of Ramsey interferometry~\cite{Knap2013}, which allows the extraction of dynamical Green's functions.

{\em Conclusions.--}
In this work, we demonstrated that reading out the mean chiral displacement of localized excitations probes the topological properties of chiral 1D many-body systems and signals the presence of symmetry-broken phases.
Furthermore, we presented a readily-feasible experimental setup where it is straightforward to realize and detect interaction-driven transitions to both symmetry-protected chiral order and symmetry-broken long-range order.
It would be interesting to achieve a deeper analytic understanding of the connected part $\xi$ of the MCD for other models, which could lead to a novel way of characterizing phases featuring spontaneous symmetry-breaking.
In particular, it is open for investigation if the MCD can distinguish all phases of even richer models.
This could be tested by including long-range hopping processes to the kinetic part of the Hamiltonian~\cite{Maffei2018} and more exotic four-body interactions~\cite{Kitaev2009}.
Further intriguing perspectives of this work are a generalization of the MCD to higher dimensions or to other symmetry-protected topological phases, and an investigation of the effects of disorder, losses and temperature on top of interactions.

\vspace{5mm}

{\em Acknowledgments.--}
The authors wish to thank M. Aidelsburger, M. Burrello, A. Dauphin, P. van Dongen, V. Gurarie, S. Manmana, S. Paeckel and P. Schmoll for inspiring and enlightening discussions.
A.H. is thankful for the financial support from the MAINZ Graduate School of Excellence, the Max Planck Graduate Center and from the COST AtomQT action CA16221.
A.H. and M.R. acknowledge support from the Deutsche Forschungsgesellschaft (DFG) through the grant OSCAR 277810020 (RI 2345/2-1).
M.R. acknowledges support from the Alexander von Humboldt foundation, the EU through the project PASQUANS and the kind hospitality of the BEC Center in Trento, where part of the manuscript writing was conducted.
P.M. acknowledges support by the ``Ram\'on y Cajal" program, the Spanish MINECO (FIS2017-84114-C2-1-P), and EU FEDER QuantumCat.
The MPS simulations were run on the Mogon cluster of the Johannes Gutenberg-Universit\"at (made available by the CSM and AHRP), with a code based on a flexible Abelian Symmetric Tensor Networks Library, developed in collaboration with the group of S. Montangero at the University of Ulm (now moved to Padua).

\bibliography{biblio}

\onecolumngrid
\begin{center}
\newpage\textbf{
\large APPENDIX}\\[4mm]

\end{center}
\setcounter{equation}{0}
\setcounter{section}{0}
\makeatletter
\renewcommand{\theequation}{S.\arabic{equation}}
\renewcommand{\thesection}{S.\arabic{section}}

The additional material is organized as follows.
In Section A we present a detailed derivation of the mean chiral displacement (MCD) in interacting systems. In particular, we prove that the disconnected (one-body) part is actually quantized to the many-body winding number $\gamma$ for a half-filled (noninteracting) Fermi sea, while at arbitrary fillings it presents time-damped oscillations around the same invariant, so that it converges to $\gamma$ in the long time limit.
Incidentally, we show that this provides an exact real-space reformulation of the many-body winding number.
In Section B we provide details on the derivation of the effective spin model presented in the main text.
In Section C we elucidate on how the long-ranged Hamiltonian $\MH_{\rm lr}$ presented in the text is directly connected to a particular limit of the Creutz model.
We conclude by discussing the exact simulations of quadratic Hamiltonians in Section D, and the MPS simulations of interacting systems in Section E.

Let us begin by presenting some definitions, which we use throughout all sections.
In this manuscript we focus specifically on chiral symmetric Hamiltonians, i.e., those which commute with an \emph{anti-unitary} operator $\widehat{\Gamma}$, acting \emph{locally} on the fermionic operators $c_{x,\tau,s}$ as $\widehat{\Gamma} c_{x,\tau,s}^{} \widehat{\Gamma}^{-1} = \Gamma_{\tau s,\tau's'} c_{x,\tau',s'}^\dagger$. Here $\Gamma$ is an even-dimensional \emph{unitary} matrix squaring to the identity, $\Gamma^2 = \mathbb1$.
As discussed in detail in Refs.~\cite{Gurarie2011,Manmana2012}, their topological properties are captured by the many-body invariant
\begin{align}\label{eq:winding_gurarie_SM}
    \gamma = {\rm tr}\int \frac{\rd k}{4\pi\ri}\,\Gamma g^{-1} \partial_k g\,,
\end{align}
where $g(k)=G(k,\omega=0)$ is the zero-frequency component of the imaginary-time Green's function.
Such an object is not straightforward to measure in common experiments, highlighting the importance of alternative formulations of the same object.

Our aim is to study the connection between the invariant $\gamma$ in Eq.~\eqref{eq:winding_gurarie_SM} and the {\it mean chiral displacement} (MCD),
\begin{align}\label{eq:mcd_definition_SM}
    \mathcal C(t) =
        \sum_{\tau,s}
        \Big\langle
            c^{}_{0,\tau,s} \re^{\ri\MH t/\hbar} \GX \re^{-\ri\MH t/\hbar} c^\dag_{0,\tau,s}
        \Big\rangle
        \,.
\end{align}
The MCD is therefore defined as the time-dependent expectation value of the {\it chiral displacement} operator $\Gamma X$,
\begin{equation}
\Gamma X \equiv \sum_x x \sum_{\tau,\tau',s,s'} c^\dag_{x,\tau,s} \Gamma_{\tau s,\tau' s'} c^{}_{x,\tau',s'}\,,
\end{equation}
(with $c_{x,\tau,s}$ destruction operators acting on the unit-cell at position $x$, sublattice $\tau$ and spin $s$) over a set of excitations generated by all the operators $\{c^\dag_{0,\tau,s}\}$ which add a particle on the central unit-cell $x=0$ at time $t=0$ over the ground state of the system.
To make a concrete example, for an SSH chain populated by spin 1/2 particles (the {\it Peierls-Hubbard} model $\mathcal{H}_{\rm sr}$ introduced in the main text), such a set of excitations could be generated by acting with the four operators $\{c^\dagger_{0,A,\up},c^\dagger_{0,A,\down}, c^\dagger_{0,B,\up},c^\dagger_{0,B,\down}\}$.
Identical results for $\mathcal{C}(t)$ are actually obtained choosing any linear combination of those which forms a complete and orthonormal basis of excitations generated at the central unit-cell. The index $i$ runs from 1 to the ``internal dimension'' of a unit cell.

Moreover, chiral symmetry ensures that around half-filling it is possible to extract the same information (up to a global sign) by removing particles at time $t=0$, instead of adding them. For example, we may write
\begin{equation}
\mathcal C(t) = - \sum_{\tau,s}\langle c^\dag_{0,\tau,s} \re^{\ri\MH t/\hbar} \GX \re^{-\ri\MH t/\hbar} c^{}_{0,\tau,s}\rangle,
\end{equation}
where the sum runs over localized hole excitations (in the specific case of $\mathcal{H}_{\rm sr}$, these excitations may be generated for example by $\{c^{}_{0,A,\up},c^{}_{0,A,\down}, c^{}_{0,B,\up},c^{}_{0,B,\down}\}$).

The most general form of the many-body MCD is therefore given explicitly by
\begin{align}
    \C(t) = \pm \sum_i\langle \psi^{}_i e^{\ri \MH t/\hbar} \GX e^{-\ri \MH t/\hbar} \psi^{\dag}_i \rangle\,,
\end{align}
with $\pm$ depending on whether the operators $\psi^\dag_i$ create particles (+), or removes them (-), and the sum runs over a complete orthonormal set of creation/destruction operators acting on a given unit cell sufficiently far away from the boundaries of the chain (to avoid scattering from the edges for times $t<T$).

\section{A: Decomposition of the mean chiral displacement through contractions}
We use the following contractions to decompose the correlator of interest, i.e., the MCD
\begin{align}
    \C(t) &=
    \xi(t)
    +
    \sum_x x
    \left\{
    \sum_{\tau,\tau',\tau'',s,s',s''}
    	\Big\langle
    		c\tikzmark{0}^{}_{0,\tau,s} c\tikzmark{00}^\dag_{x,\tau',s'}(t)
    		c\tikzmark{1}^{\vphantom\dag}_{x,\tau'',s''}(t)c\tikzmark{11}^{\dag}_{0,\tau,s}
	\Big\rangle \Gamma_{\tau's',\tau''s''}
    \right\}
    \\\nonumber
    &=
    \xi(t)
    +
    \sum_x x
    \left\{
    \sum_{\tau,\tau',\tau'',s,s',s''}
    \left[
    \langle c_{0,\tau,s}^{\vphantom\dag}c_{0,\tau,s}^{\dag}\rangle    \langle c^\dag_{x,\tau',s'}(t)c^{\vphantom\dag}_{x,\tau'',s''}(t)\rangle
	+
    \langle c_{0,\tau,s}^{\vphantom\dag}c^\dag_{x,\tau',s'}(t)\rangle 	\langle c^{\vphantom\dag}_{x,\tau'',s''}(t)c_{0,\tau,s}^{\dag}\rangle
    \right]
    \Gamma_{\tau's',\tau''s''}
    \right\}
\end{align}
in which $\xi(t)$ denotes the many-body (connected) part of the MCD.
\begin{tikzpicture}[remember picture, overlay]
    \coordinate (1) at ([yshift=3.5ex]pic cs:0);
    \coordinate (2) at ([yshift=3.5ex]pic cs:00);
    \draw ([yshift=2.5ex]pic cs:0) -- (1) -- (2) -- ([yshift=2.5ex]pic cs:00);
    \coordinate (1) at ([yshift=3.5ex]pic cs:1);
    \coordinate (2) at ([yshift=3.5ex]pic cs:11);
    \draw ([yshift=2.5ex]pic cs:1) -- (1) -- (2) -- ([yshift=2.5ex]pic cs:11);

    \coordinate (1) at ([yshift=-3ex]pic cs:0);
    \coordinate (2) at ([yshift=-3ex]pic cs:11);
    \draw ([yshift=-1ex]pic cs:0) -- (1) -- (2) -- ([yshift=-1ex]pic cs:11);
    \coordinate (1) at ([yshift=-2ex]pic cs:00);
    \coordinate (2) at ([yshift=-2ex]pic cs:1);
    \draw ([yshift=-1ex]pic cs:00) -- (1) -- (2) -- ([yshift=-1ex]pic cs:1);
\end{tikzpicture}
The one-body (disconnected) part of the MCD is therefore defined by $\C_d = \C - \xi$.
Note that, for a noninteracting system, the contractions correspond to the application of Wick's theorem and, obviously, $\xi=0$.
In the above decomposition, we made use of the particle number conservation to neglect expectation values with unequal number of creation and annihilation operators.
Furthermore, since $t>0$, we can readily identify the single-particle (retarded) Green's functions
$\ri G_{\tau s,\tau' s'}(x,t) \equiv \langle c^{\vphantom\dag}_{x,\tau,s\vphantom'}(t)c^\dag_{0,\tau',s'}(0)\rangle$ and
$\langle c^{\vphantom\dag}_{0,\tau',s'}(0)c^\dag_{x,\tau,s\vphantom'}(t)\rangle = \langle c^{}_{x,\tau,s\vphantom'}(t)c^\dag_{0,\tau',s'}(0)\rangle^* = -\ri G_{\tau s, \tau's'}^*(x,t)$.
Using the definition of the chiral displacement and the matrix definitions of both $G$ and $\Gamma$ (their indices being spin and sublattice ones), we obtain:
\begin{align}
    \C = \xi(t) + \C_d(t) =
    	\xi(t) + \sum_{\tau,s}\braket{1-n_{0,\tau,s}}\braket{\GX} + \sum_x x \cdot \tr \left(G^\dag\Gamma G\right)(x,t)\,,
\end{align}
which is Eq.~\eqref{eq:MCD_wicks_theorem} of the main text.
It is important to notice that the first sum does not depend on time, because the average $\langle\GX\rangle$ is intended over the state before the perturbation, which is assumed to be an eigenstate of $\MH$ (either the true vacuum or the half-filled system, in our work here).

Let us now focus on the time-dependent one-body part of the correlator $\C$, i.e. $\DC_d(t) \equiv \sum_x x \cdot\tr\left(G^\dag\Gamma G\right)(x,t)$.
Using the Fourier transform $G(x,t) = (2\pi)^{-1}\int {\rm d}k\,e^{i kx}G(k,t) = (2\pi)^{-2} \int {\rm d}k\, \int {\rm d}\omega\, e^{i (k x -\omega t)} G(k,\omega)$ and performing an integration by parts (to substitute $x\rightarrow \ri\partial_k$) we arrive at
\begin{align}
\Delta\C_d(t)=\C_d(t)-\C_d(0)=&
\frac{\ri }{2\pi}\int{\rm d}k \  \tr \left[G^\dagger(k,t) \, \Gamma \partial_k G(k,t)\right]
\label{eq:DC_k_t} \\
= &
\frac{\ri }{(2\pi)^3}\int{\rm d}k\,{\rm d}\omega\,{\rm d}\omegaT\  \tr \left[G^\dagger(k,\omega) \, \Gamma \partial_k G(k,\omegaT) \right] e^{i(\omega-\omegaT)t} \,.
\label{eq:DC_k_omega}
\end{align}
The identity $\partial_k \left( G G^{-1} \right)=0$ allows one to write  $\partial_k G(k,\omegaT) = - G(k,\omegaT) \partial_k G^{-1}(k,\omegaT) \, G(k,\omegaT)$.
Upon commuting $\Gamma G(k,\omegaT) = - G(k,-\omegaT) \Gamma$ one finds
\begin{align}
\Delta\C_d(t)= & \frac{\ri }{(2\pi)^3}\int{\rm d}k\,{\rm d}\omega\,{\rm d}\omegaT\
\tr \left[G^\dagger(k,\omega) G(k,-\omegaT)\, \Gamma \, \partial_k G^{-1}(k,\omegaT) \, G(k,\omegaT) \right]e^{i(\omega-\omegaT)t}\,.
\end{align}

To proceed further, we consider now a non-interacting Fermi sea in a lattice with unit-cells containing two sites only (like the famous SSH model, for example).
The lattice filling is controlled by the Fermi energy $\varepsilon_F$, and the Green's function may be written as
\begin{equation}\label{G0}
G(k,\omega) = \frac{\theta(H_0(k)-\varepsilon_F)}{\omega-H_0(k)+i 0^+} + \frac{\theta(\varepsilon_F-H_0(k))}{\omega-H_0(k)-i 0^+}\, .
\end{equation}
Before proceeding with the calculation of $\Delta\C_d(t)$, let us recall that in an appropriate basis, the Hamiltonian may be cast in a completely off-diagonal form to anti-commute with the chiral operator $\Gamma=\sigma_z$:
\begin{align}\label{eq:non-interacting-Hamiltonian}
    H_0(k) =
    h_x(k)\sigma_x + h_y(k)\sigma_y
    =
    \varepsilon_k\,\vec n_k\cdot\vec\sigma\,
    =
    \varepsilon_k
    \begin{pmatrix}
        0 & w_k \\
        w^*_k & 0
    \end{pmatrix}
    \,,\quad
    w_k = \frac{h_x(k) - \ri h_y(k)}{\varepsilon_k}
    \,,
\end{align}
with  $\varepsilon_k=\sqrt{|h_x(k)|^2+|h_y(k)|^2}$, and $\vec n$ is a vector of unit norm.
In this case $g(k) = -H_0^{-1}(k) = -H_0(k)/\varepsilon_k^2$ and thus the integral leading to the invariant in Eq.~\eqref{eq:winding_gurarie_SM} simplifies to
\begin{align}
    \gamma
= \frac1{4\pi\ri}{\rm tr}\int {\rd k}\,\Gamma g^{-1} \partial_k g
= \frac{1}{4\pi\ri}\int \rd k \left(w_k^{\vphantom*}\partial_k^{\vphantom*}w^*_k - w^*_k\partial_k^{\vphantom*} w_k^{\vphantom*} \right)
= \frac{1}{2\pi\ri}\int \rd k\,w^{\vphantom*}_k\partial_kw^*_k\,,
\end{align}
since $|w_k|=1$, so that $\partial_k^{\vphantom*}(w_k^{\vphantom*}w_k^*) = 0$.
%

\subsection{A.1: Adding one particle on top of the vacuum}

When a particle is added to a completely empty lattice, as considered originally in Refs.~\cite{Cardano2017, Maffei2018, Meier2018}, the chemical potential coincides with the bottom of the band, so that the Green's function reads simply $G(k,\omega) = 1/(\omega-H_0(k)+i 0^+ )$, and all its poles are in the lower half of the complex plane.
This allows for the simplification $\partial_k G^{-1}(k,\omega) = -\partial_k H_0(k) \equiv -H'_0$.
Making use of Cauchy's residue theorem and closing the integration contour in the lower (upper) semi-plane for $G$ ($G^\dagger$), one finds
\begin{align} \label{eq:disconnected-MCD_SM}
\Delta\C_d(t)
& = \frac{\ri}{(2\pi)^3}\int{\rm d}k\,\tr \left\{
\left[\int{\rm d}\omega\,G^\dagger(k,\omega) e^{\ri \omega t}\right]
\left[\int{\rm d}\omegaT\, G(k,-\omegaT)\, \Gamma \, H'_0\, G(k,\omegaT) e^{-\ri\omegaT t}\right]\right\}\\\nonumber
& = \frac{\ri}{(2\pi)^3}\int{\rm d}k\,\tr \left\{
\left[(2\pi \ri) e^{\ri H_0(k) t}\right]
\left[(2\pi\ri)\left(\frac{1}{2H_0(k)}\Gamma\, H'_0\, e^{- \ri H_0(k) t}-e^{\ri H_0(k) t}\frac{1}{2H_0(k)}\Gamma\, H'_0\right)\right]\right\}\nonumber\\\nonumber
& = \frac{1}{4\pi {\rm i}}\int{\rm d}k\,\tr \left\{
\left[1- e^{\ri 2H_0(k) t}\right]
\frac{1}{H_0(k)}\Gamma\, H'_0\right\}\\\nonumber
& =\gamma -\frac{1}{4\pi {\rm i}}\int{\rm d}k\,\tr \left\{
e^{\ri 2H_0(k) t}\,\Gamma g^{-1}(k)
 \partial_k [g(k)]\right\}\,.
\end{align}
In the last step, we used $[\Gamma,1/H_0(k)]=0$, then $H_0=-1/g$, and finally $g\partial_k(g^{-1}) = \partial_k \ln(g^{-1}) = - g^{-1} \partial_k g$.
Due to the presence of the time $t$ inside the phase factor $e^{\ri 2H_0(k) t}$, the integrand of the last term on the r.h.s. changes sign very rapidly when $t\rightarrow\infty$, so that its integral is increasingly damped, and as time grows the whole expression converges smoothly to the chiral invariant $\gamma$ introduced in Eq.~\eqref{eq:winding_gurarie_SM}. In terms of the matrix element $w_k$ an the energy eigenvalue $\varepsilon_k$ introduced in Eq.~\eqref{eq:non-interacting-Hamiltonian}, the matrix expression Eq.~\eqref{eq:disconnected-MCD_SM} simplifies notably to the scalar result
\begin{align}
\Delta\C_d(t)= \label{MCD_single_particle_SM}
  \frac{1}{2\pi\ri}\int \rd k \left[1-\cos(2 \varepsilon_k t)\right] w_k\partial_k w_k^*
= \gamma-\frac{1}{2\pi\ri}\int \rd k\,\cos(2 \varepsilon_k t)\, w_k\partial_k w_k^*\,,
\end{align}
which shows more explicitly how the result is the chiral winding $\gamma$, plus a damped oscillatory term (guaranteed to be purely real). The latter result, derived here by means of a many-body formalism, coincides exactly with the one found in Refs.~\cite{Cardano2017}.

A lengthy but straightforward calculation (in analogy to the one presented in Ref.~\cite{Cardano2017}) allows one to show that for a system with unit-cells containing two sites only $\C(t)=2\C_{1/2}(t)$, where $\C_{1/2}(t)\equiv \langle\Gamma X\rangle_{\psi}$ is the MCD computed over a single excitation, instead of over a collection of them as in Eq.~\eqref{eq:mcd_definition_SM} (provided of course the system has global  spin-rotation invariance; else, a sum on spins should also be performed). This convenient feature was used repeatedly in the main text.


\subsection{A.2: Adding one particle to a half-filled Fermi sea}
As the lattice is increasingly filled, the poles of $G$ ($G^\dagger$) which are below the Fermi momentum move to the upper (lower) plane, so that they exit the integration contours defined above and do not contribute to the final result:
Let us for simplicity consider a scalar $H_0$. For a given $k$, the first energy integral in (S.14) is non-zero only if $\omega_0=H_0(k)>0$, and gives a result proportional to $e^{i\omega_0t}$.
If this is the case, in the second integral only the pole at $\omega=+\omega_0$ contributes, and gives a contribution $e^{-i\omega_0t}$. The ``co-rotating term'' (which would produce a term $e^{+i\omega_0t}$) instead vanishes. As such, the result is non-oscillatory.
This reasoning can be straightforwardly extended to $H_0$ being a matrix.

In conclusion, the MCD of a half-filled Fermi sea leads to the very simple and appealing result
\begin{align}\label{MCD_groundstate_SM}
\Delta\C_d(t)
=&\frac{1}{4\pi {\rm i}}\int{\rm d}k\,\tr \left\{
\Gamma g^{-1}(k)
 \partial_k [g(k)]\right\} = \gamma\,,
\end{align}
directly quantized to the chiral invariant (i.e., without any oscillatory term).
The latter constitutes an analytical proof that the disconnected time-dependent part $\DC_d=\sum_x x\,\tr(G^\dag\Gamma G)(x,t)$ of the MCD, evaluated on a half-filled non-interacting system, is the real-space and real-time equivalent of the many-body winding defined in Eq.~\eqref{eq:winding_gurarie_SM}.
This feature was also explicitly demonstrated numerically in Fig.~\ref{fig:recipe}(c) of the main text.

Furthermore, a lengthy but otherwise straightforward calculation shows that the oscillatory term in Eqs.~\eqref{eq:disconnected-MCD_SM}-\eqref{MCD_single_particle_SM} in the previous subsection is caused by inter-band scattering processes between conduction and valence bands which are Fermi-blocked at half-filling.
Interactions re-enable such scattering processes, introducing again oscillatory contributions on top of the quantized winding number.

\section{B: Derivation of the effective spin model}
We start by introducing the link-basis $D_{x,s}=(\ell_{x,+,s},\ \ell_{x,-,s})^T$
defined via
\begin{align}
    \ell_{x,+,s} = (c_{x+1,A,s} + c_{x,B,s})/\sqrt2
    \qquad
    \ell_{x,-,s}=(c_{x+1,A,s} - c_{x,B,s})/\sqrt2
\end{align}
This self-adjoint map transforms the tight-binding terms of the spinful SSH chain as
\begin{align}
   c^\dag_{x+1,A,s} c^{\vphantom\dag}_{y,B,s} =
   \frac12 D_{x,s}^\dag \left(\sigma_z - \ri \sigma_y\right)D_{y,s}^{\vphantom\dag}
\end{align}
with the Pauli matrices being \mbox{$\sigma_i$, $i\in\{x,y,z\}$}, and densities read
\begin{align}
   n_{x'+1,A,s}
   =
   \frac12
   D_{x',s}^{\dag}
   (\mathbb1 + \sigma_x)
   D_{x',s}^{\vphantom\dag}
   \qquad
   n_{x',B,s}=\frac12
   D_{x',s}^{\dag}
   (\mathbb1 - \sigma_x)
   D_{x',s}^{\vphantom\dag}\,.
\end{align}
The bare kinetic term transforms to
\begin{align}
   \MH_0 &=
   \frac J2\sum_{x',s}
   \left\{
   D_{x'-1,s}^{\dag}
   \left(\sigma_z - \ri \sigma_y\right)
   D_{x',s}^{\vphantom\dag}
   +
   \hc
   \right\}
   +
   J'\sum_{x',s}
   D^{\dag}_{x',s}
   \sigma_z
   D^{\vphantom\dag}_{x',s}\,.
\end{align}
If we define the local pseudospin operators as
\begin{align}
    S^i_{x',s} = \frac12 D^\dag_{x',s}\sigma_i D^{}_{x',s}
    \quad\text{with}\quad i\in\{x,y,z\}
    \quad\text{and}\quad S^0_{x',s} = \frac12 D^\dag_{x',s}D^{}_{x',s} = \frac12(n_{x'+1,A,s}+n_{x',B,s})
\end{align}
it becomes clear that the term dictated by $J'$ commutes with all local densities $S^0_{x,s}$, whereas $J$-contributions exchange particles between adjacent links and thus do not commute with $S^0_{x,s}$.
Interactions result (up to irrelevant chemical potentials and boundary terms) in
\begin{align}
    \MH_\perp = 2U_\perp\sum_{x'}\left(S^0_{x',\up} S^0_{x',\down} + S^x_{x',\up}S^x_{x',\down}\right)
    \qquad
    \MH_\parallel = U_\parallel\sum_{x',s}\left(S^0_{x'-1,s} + S^x_{x'-1,s}\right)\left(S^0_{x',s} - S^x_{x',s}\right)\,.
\end{align}
We now assume a fully translational-invariant system at half filling, which implies setting the link density $S^0_{x,s}=1/2$ at every link, and arrive thus at the effective model presented in Eq.~\eqref{eq:hamiltonian_pseudospins} of the main text.

\section{C: Relation to the Creutz model}
\begin{figure}[ht]
    \includegraphics{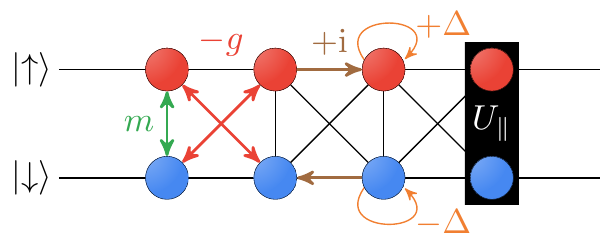}
    \caption{The interacting Creutz model. We represent hopping processes by arrows indicating the direction of movement with corresponding colored amplitude. Interactions are represented by colored boxes.}
    \label{fig:creutz_model_picture}
\end{figure}

\begin{figure}[ht]
    \includegraphics{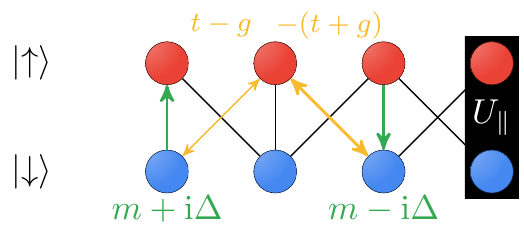}
    \caption{The interacting rotated Creutz model at $\chi=\pi$. We represent hopping processes by arrows indicating the direction of movement with corresponding colored amplitude. Interactions are represented by colored boxes.}
    \label{fig:rotated_creutz_model}
\end{figure}

The Creutz model was first introduced in Ref.~\cite{Creutz1999}, and more recently studied in detail in Refs.~\cite{Juenemann2017,Tirrito2019}.
Its kinetic part is defined as
\begin{align}
    \MH_C = \sum_{i=1}^L\left[
    c^\dag_i
    \left(t\re^{\ri\frac\chi2\sigma_z}-g\sigma_x\right)
    c^{\phantom\dag}_{i+1} + \hc
    \right]
    +
    c^\dag_i
    \left(m\sigma_x + \frac\Delta2\sigma_z\right)
    c^{\phantom\dag}_i\,.
\end{align}
in which $c = (c_{\uparrow}, c_{\downarrow})^T$ is a two-component spinor of fermionic annihilation operators.
In addition to the kinetic part, we consider the interaction
$H_{\rm int}=U_\parallel \sum_i n_{\uparrow,i} n_{\downarrow,i}$ resulting in a lattice according to Fig.~\ref{fig:creutz_model_picture}.
The chiral symmetry operator in this basis takes the form $\widetilde\Gamma = \sigma_y$.
It is more convenient to work in a basis in which $\Gamma = \sigma_z$, which can be achieved by rotating the former coordinate system by an angle $\pi/2$ around the $\sigma_x$ axis.
Then, $\sigma_y\rightarrow\sigma_z$, $\sigma_x\rightarrow\sigma_x$ and $\sigma_z\rightarrow-\sigma_y$, which leads us to a rotated Creutz model
\begin{align}
    \MH_{\rm RC} = \sum_{i=1}^L\left[
    c^\dag_i
    \left(t\re^{-\ri\frac\chi2\sigma_y}-g\sigma_x\right)
    c^{\phantom\dag}_{i+1} + \hc
    \right]
    +
    c^\dag_i
    \left(m\sigma_x - \frac\Delta2\sigma_y\right)
    c^{\phantom\dag}_i\,.
\end{align}
The new lattice for $\chi=\pi$ displayed in Fig.~\ref{fig:rotated_creutz_model} is slightly simpler, in the sense that the horizontal hoppings have been replaced by complex spin-mixing terms.

\begin{figure}[ht]
    \includegraphics{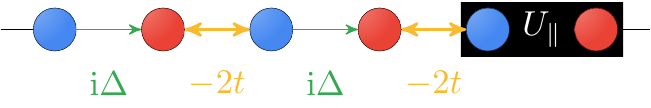}
    \caption{The tilted lattice of Fig.~\ref{fig:rotated_creutz_model} at the fine-tuned point $m=0$ and $g=t=\chi/\pi$.}
    \label{fig:rotated_creutz_model_ssh}
\end{figure}

At the fine-tuned point $m=0$ and $t=g=\chi/\pi$ it is possible to find a connection to the spinless SSH chain:
By tilting (and stretching) the lattice of Fig.~\ref{fig:rotated_creutz_model} at $g=t$, it becomes apparent that we are dealing with a spinless SSH chain in which $\uparrow$ and $\downarrow$ play the role of $A$ and $B$ sublattices, subject to a density-density interaction which acts within the unit-cells (see Fig.~\ref{fig:rotated_creutz_model_ssh}).
Indeed, this fine-tuned point results in the nontrivial phase diagram studied in Ref.~\cite{Juenemann2017}, which was later studied in the context of Wilson-Hubbard matter in high energy physics~\cite{Bermudez2018,Tirrito2019,Kuno2019}.
We thus arrive at the Hamiltonian $\MH_{\rm lr}$ studied in the main text.

\section{D: Time-evolution of quadratic Hamiltonians}
Here we focus on generic non-interacting systems, describe in detail the modelling of single-particle excitations and their time-evolution by exact diagonalization of quadratic Hamiltonians.
Thus, we assume a generic Hamiltonian which can be cast into the form
\begin{align}
    \MH_0 = \sum_{i,j=1}^L h_{ij}c^\dag_i c^{\phantom\dag}_j\,,
\end{align}
in which $h_{ij}$ is a $L\times L$ matrix and $c_j$ are fermionic annihilation operators at lattice position $j$.
We choose to rotate into the eigenbasis of $\MH_0$, i.e.
\begin{align}
    \MH_0 = \sum_k \varepsilon_k d_k^\dag d_k^{\phantom\dag}
\end{align}
is the diagonalized Hamiltonian of eigenmodes $d_k = \sum_jU_{kj} c_j$ with eigenenergies $\varepsilon_k$, which we choose to sort by magnitude $\varepsilon_k\leq\varepsilon_{k+1}$.
In this basis, the ground state is of the form
\begin{align}
    \Ket{\Psi} = \prod_{n\leq N}d_n^\dag\Ket{0}\,,
\end{align}
in which $N$ denotes the total number of fermions.
Applying the definition of the ground state and rotating the basis, we find the single-particle Green's function
\begin{align}
    G^0_{ij} =\Braket{\Psi|c^\dag_i c_j|\Psi}
    = \sum_{p\leq N}U^T_{ip}U^*_{pj}\,.
\end{align}
We want to study the dynamics of a generic excitation of holes (or particles).
In particular,
\begin{align}
    \Ket{\psi^{+}} = \frac1{\sqrt \Omega}\sum_x Q^{\phantom\dag}_x c^{\dag}_x\Ket{\Psi}
    \qquad
    \Ket{\psi^{-}} = \frac1{\sqrt \Omega}\sum_x Q^{\phantom\dag}_x c^{\phantom\dag}_x\Ket{\Psi}\,,
\end{align}
where we introduced an (unnormalized) vector $Q\in\mathbb R^L$ which is modeling the annihilation (creation) of local modes $c^{(\dag)}_x$.
For example, a single excitation at position $L/2$ would imply a $Q$ with one nonzero entry $Q_{L/2}=1$.
The normalization constant $\Omega$ can be explicitly calculated from $\Braket{\psi^{\pm}|\psi^{\pm}}$ and evaluates to
\begin{align}
    \Omega = \sum_{p\in \mathcal{P^{\pm}}} \left|U_{px}Q_x\right|^2 = \sum_{p\in \mathcal{P^{\pm}}}\left|P_p\right|^2
\end{align}
with the excitation vector $P=UQ$ and the subset $\mathcal{P^+}=\{p,p>N\}$ for the creation of particles, or $\mathcal{P^-}=\{p,p\leq N\}$ for the creation of holes.
We are interested in the time-evolution of $\Ket{\psi^\pm}$, which is given after application of the time-evolution operator
\begin{align}
    A(t) = \re^{-\ri H_0 t} = \sum_k d^\dag_k \re^{-\ri \varepsilon_k t} d^{\phantom\dag}_k= \sum_k d^\dag_k A_k^{\phantom\dag} d^{\phantom\dag}_k\,.
\end{align}
If we take a momentum eigenstate of the Hamiltonian $d^\dag_k\Ket{0}$ and apply $A(t)$, we find the time-evolved state
\begin{align}
     A(t) d_k^\dag\Ket{0} = A_k^{\phantom\dag} d_k^\dag\Ket{0} = t^\dag_k\Ket{0}
\end{align}
and we note that we may interpret time-evolution as a rotation of the diagonal basis according to $t_k^{\dag} = A_k^{\phantom\dag} d_k^{\dag}$.
The resulting set of operators $t_k^\dag$ are creators of time-evolved eigenmodes at momentum $k$.
Note, that the time-evolved excitations read
\begin{align}
    \Ket{\psi^{+}(t)} = \frac1{\sqrt \Omega}\sum_{k\in\mathcal{P^{+}}} P_k t^\dag_k\Ket{\Psi (t)}
    \qquad
    \Ket{\psi^{-}(t)} = \frac1{\sqrt \Omega}\sum_{k\in\mathcal{P^{-}}} {P_k^*} t^{\phantom\dag}_k\Ket{\Psi (t)}\,
\end{align}
with the time-evolved ground state
\begin{align}
    \Ket{\Psi(t)} = \prod_{p\leq N} t_p^\dag\Ket{0}\,.
\end{align}
Now, we have all ingredients to calculate the Green's function of the time-evolved excited state
\begin{align}
    G^{\psi^{\pm}}_{ij}(t)
    =
    G_{ij}^0
    \pm
    \frac1\Omega
    \left(\sum_{p\in\mathcal{P}^{\pm}}P_p^* A^{\dag}_pU^{\phantom\dag}_{pi}\right)
    \left(\sum_{q\in\mathcal{P}^{\pm}}U^\dag_{jq}A^{\phantom\dag}_{q}P^{\phantom\dag}_q\right)
\end{align}
which can be used to compute any non-interacting expectation value (for example, its diagonal entries give the values of the local density).
We note the static part $G^0_{ij}$, which yields a potential constant (in time) in the MCD.
This constant can be amended by the quantity $\DC = \C(t) - \C(t=0^-)$, with $\C(t=0^-)$ the (static) MCD of the ground state right before the creation of the local excitation.

\begin{figure}[ht]
    \includegraphics{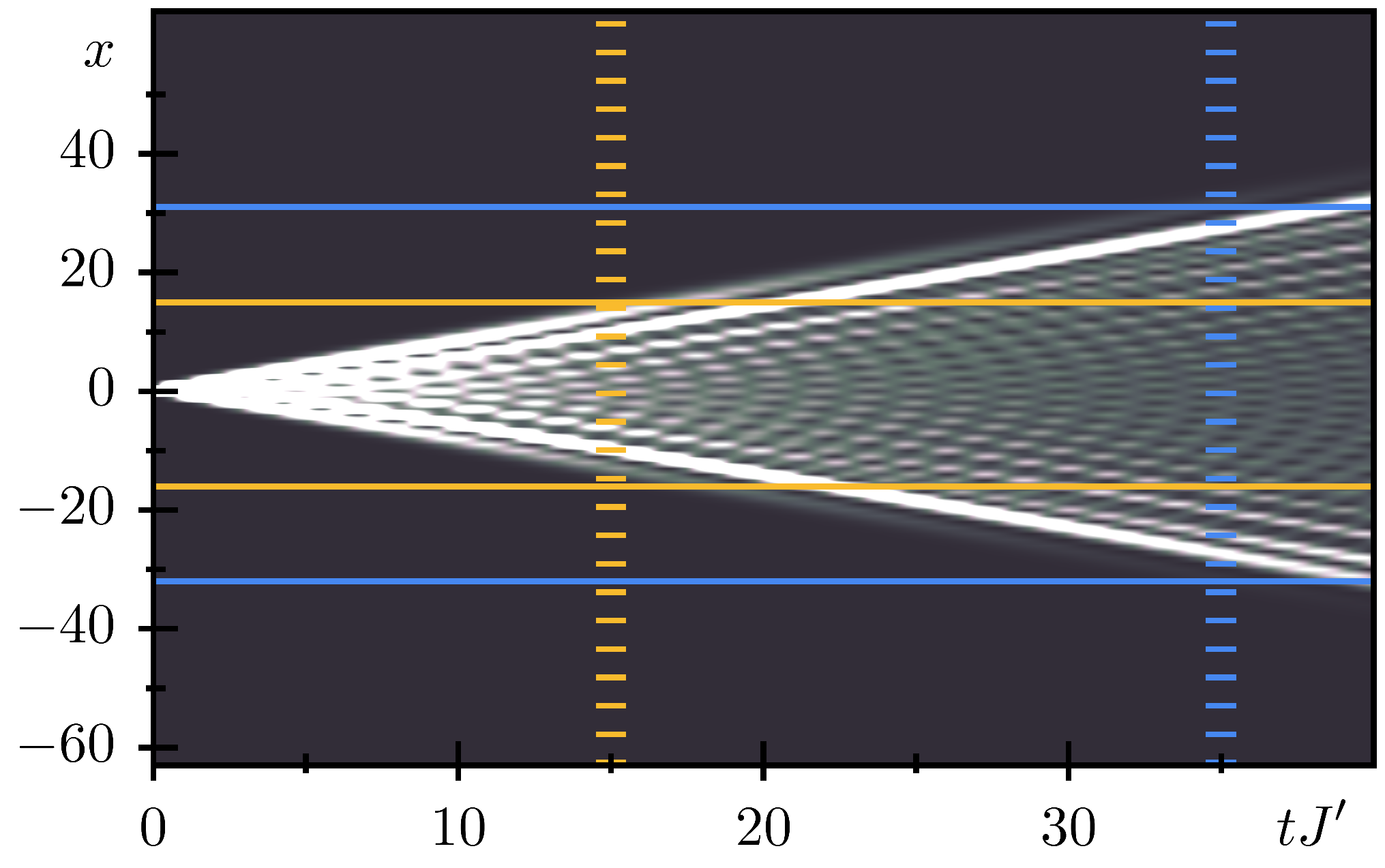}\llap{\parbox[b]{14.7cm}{\color{white}(a)\\\rule{0ex}{4.9cm}}}
    \hfill
    \includegraphics{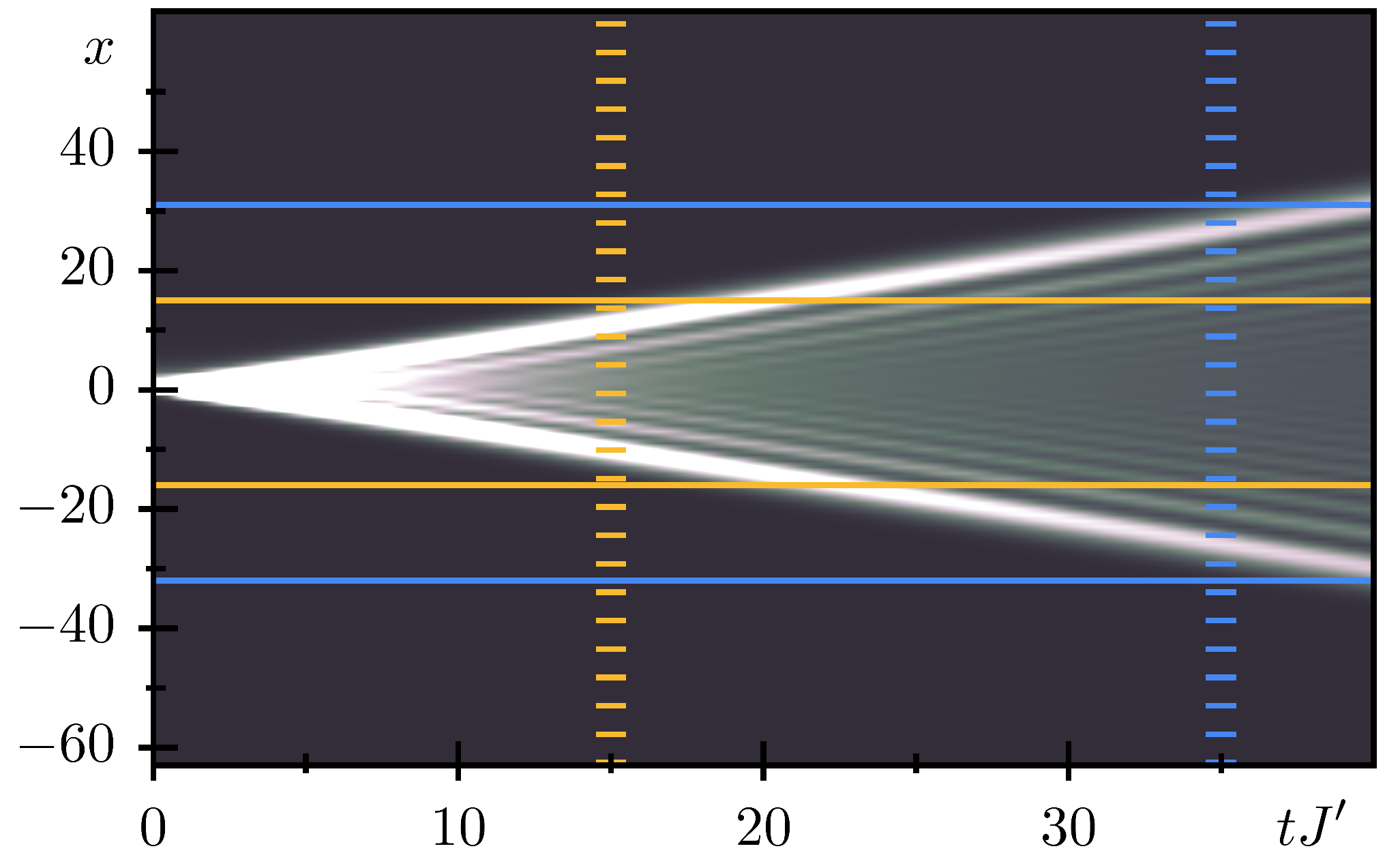}\llap{\parbox[b]{14.7cm}{\color{white}(c)\\\rule{0ex}{4.9cm}}}
    \includegraphics{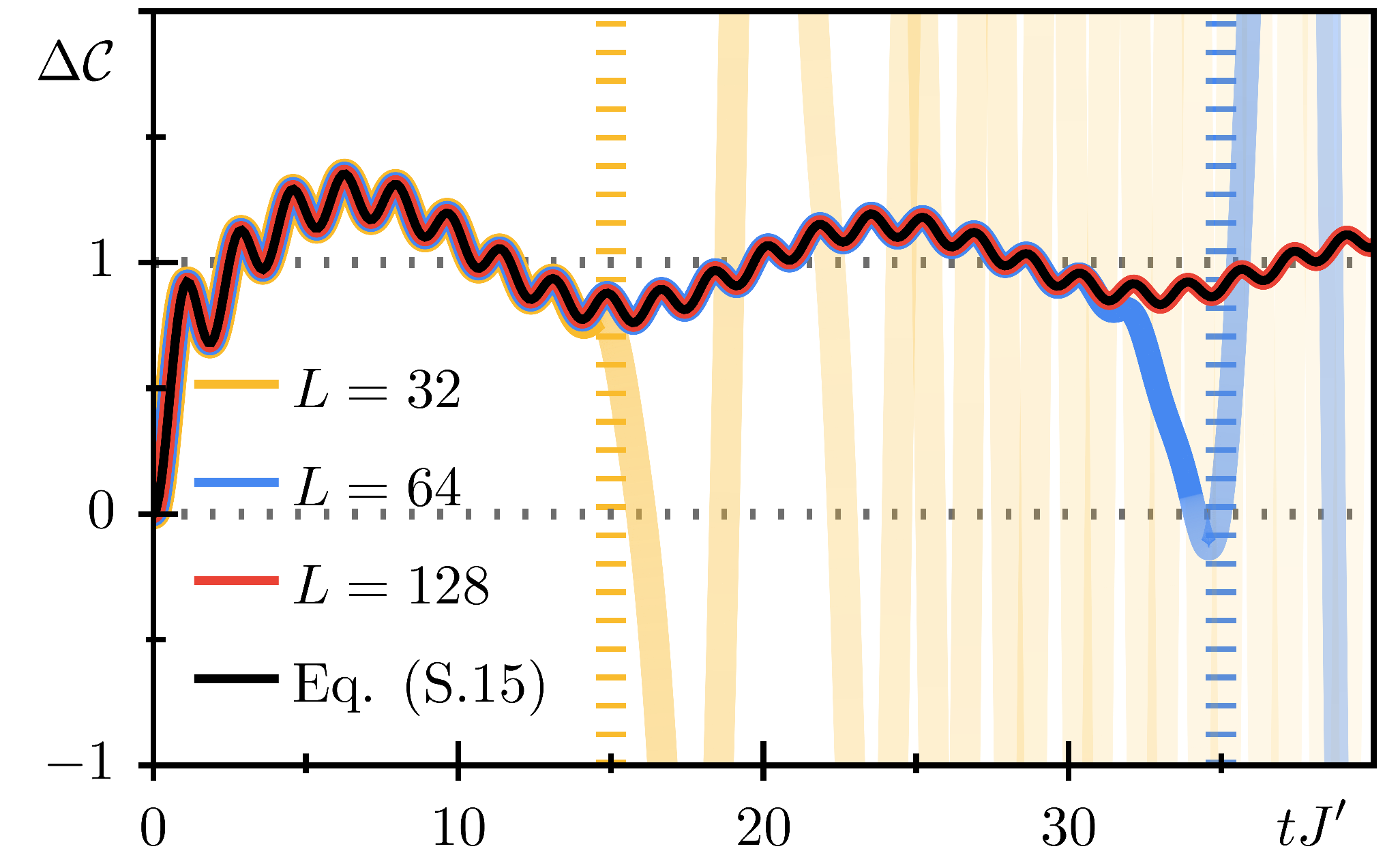}\llap{\parbox[b]{14.7cm}{(b)\\\rule{0ex}{4.6cm}}}
    \hfill
    \includegraphics{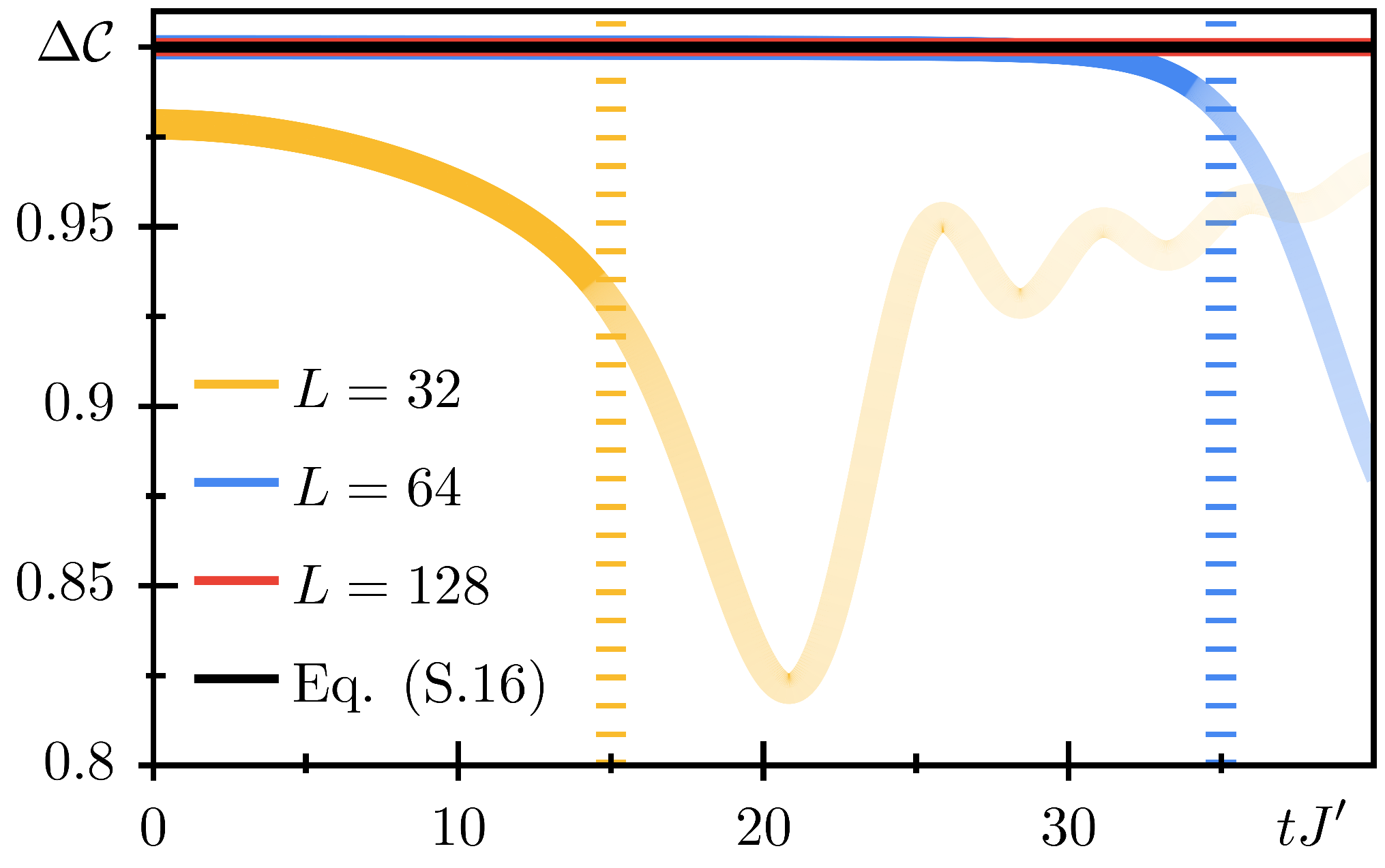}\llap{\parbox[b]{14.7cm}{(d)\\\rule{0ex}{4.6cm}}}
    \caption{Panel (a) displays the density bulk-propagation of a single-particle excitation above the vacuum for a single SSH chain at $J/J' = 0.7$ for $L=128$ unit-cells (i.e., the topological phase with winding number $\gamma=1$) and open boundary conditions. The grid lines display times at which the excitation would scatter with the boundary of smaller system sizes, which results in deviations from the fully translationally invariant solution of the MCD presented in (b).
    The black line is the analytical solution of Eq.~\eqref{MCD_single_particle_SM}, and colored lines represent numerical solutions on chains with different lenghts.
    Panel (c) depicts the density bulk-propagation of a single-particle excitation above a half-filled Fermi sea with the same system parameters as in panel (a).
    The nontrivial interference pattern visible in the vacuum case disappears in this case.
    Panel (d) displays the MCD, which becomes quantized in the thermodynamic limit in agreement with Eq.~\eqref{MCD_groundstate_SM} (shown as a black line).
    For finite systems, minute oscillations appear when the excitation scatters with the boundary.
    }
    \label{fig:fs_non-interacting}
\end{figure}

To test finite-size scaling of the MCD, in Fig.~\ref{fig:fs_non-interacting} we compare the evolution of excitations in finite chains with variable lengths to the fully translational invariant and infinite sized limits derived in Section A. As one may have naively expected, the local densities (and therefore the MCD results) are completely identical for times in which the excitation is spreading only through the bulk of the system.

\section{E: Details on the MPS simulations}
All MPS simulations of interacting systems presented here and in the main text are performed on an open boundary conditioned system with $L=32$ unit-cells.
For simplicity we took the coupling constants $J,J'$ to be real-valued, i.e. we considered BDI Hamiltonians.
In case of $\MH_{\rm lr}$, the system is fully decoupled along the spin such that the unit-cell can be reduced to that of a spinless SSH-chain (e.g., we consider only the $\up$-wire).
We consider up to $M=256$ Schmidt-values and simulate the time-evolution via the time-dependent variational principle (TDVP)~\cite{Haegeman2011}, using a timestep of at most $\Delta t=0.1J'$.

An estimate for the maximum bulk evolution time (at least in the absence of interactions) can be easily derived from the maximum group velocity of the single-particle excitations.
The dispersion of the SSH chain is given by $\varepsilon(k)=\sqrt{J^2+J'^2+2JJ'\cos(k)}$ and we find $v_{\rm max} = \max_k v(k) = \min(J,J')$ (assuming dimensionless units as in our numerical simulations).
For a localized wavepacket at the center of the chain to not reach the boundaries for a given tunneling $J'$, the maximum velocity is bounded by $v_{\rm max}/J' = \min(J/J',1) < L/(2TJ')$, which is always satisfied if we restrict the time window to $2TJ'< L$.
The wavepacket will instead hit the boundaries at times $tJ'=L/2 + nL$, $n\in\mathbb N$.
This is demonstrated explicitly in Fig. 1(b) and 1(c), where we present results for a non-interacting half-filled SSH chain.
The resulting Hamiltonian is quadratic and it can be easily diagonalized, so that the evolution shown in the Figure is {\it exact}.
On a lattice containing $L= 64$ unit cells, the wavepacket hits the boundaries first at time $tJ'\approx 32$, then its reflected, and hits the boundaries again at time $tJ'\approx 96$ [see panel (b)].
To avoid boundary effects, we restrict the time-window to $t< 10J'$, such that the propagating wavepacket is never reflected by the boundaries.

\begin{figure}[ht]
    \includegraphics{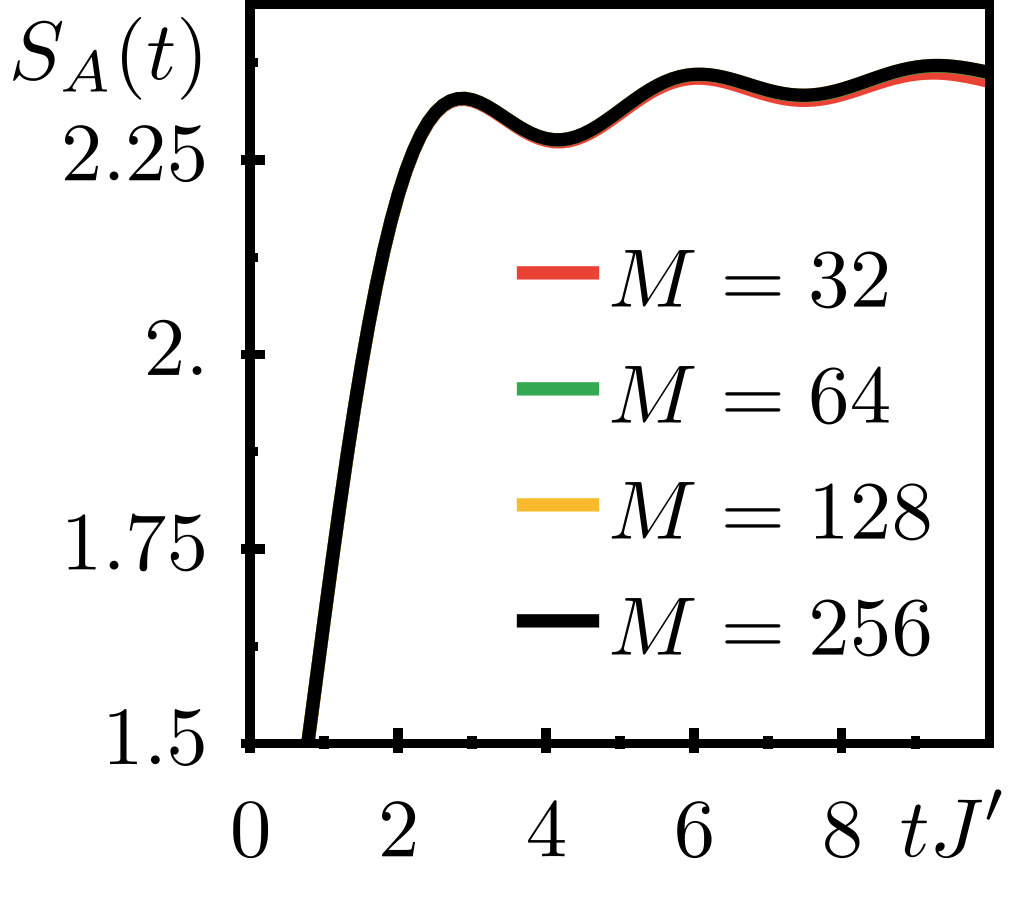}\llap{\parbox[b]{5.8cm}{(a)\\\rule{0ex}{3.3cm}}}
    \includegraphics{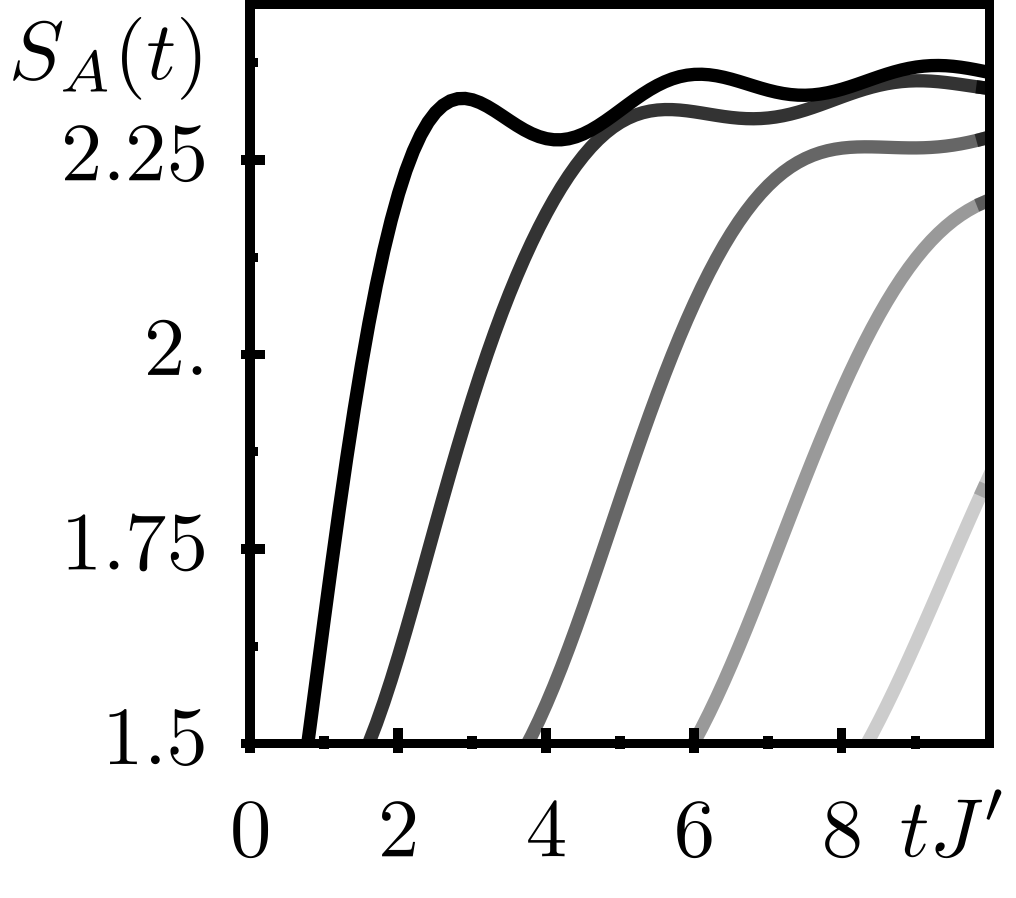}\llap{\parbox[b]{5.8cm}{(b)\\\rule{0ex}{3.3cm}}}\llap{\parbox[b]{5.8cm}{1\\\rule{0ex}{2.3cm}}}\llap{\parbox[b]{4.8cm}{2\\\rule{0ex}{2.3cm}}}\llap{\parbox[b]{3.2cm}{3\\\rule{0ex}{2.3cm}}}\llap{\parbox[b]{1.6cm}{4\\\rule{0ex}{2.3cm}}}\llap{\parbox[b]{0.9cm}{5\\\rule{0ex}{1.3cm}}}
    \includegraphics{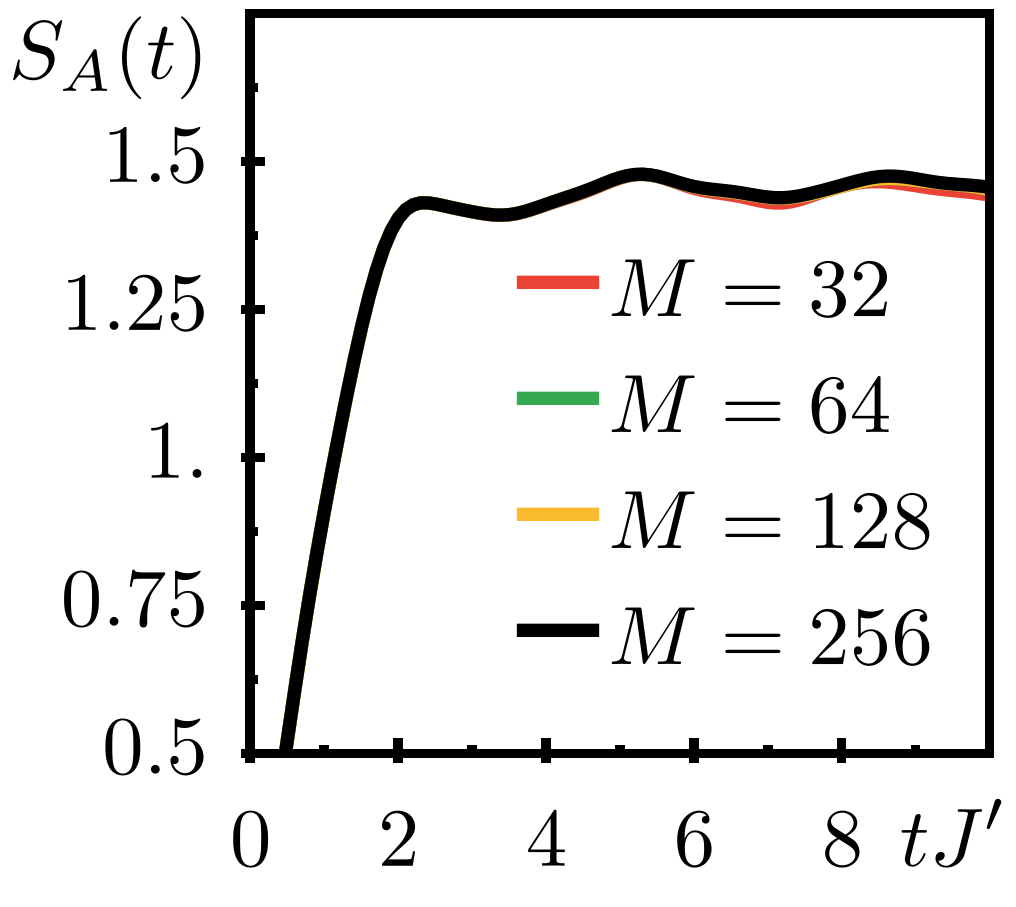}\llap{\parbox[b]{5.8cm}{(c)\\\rule{0ex}{3.3cm}}}
    \includegraphics{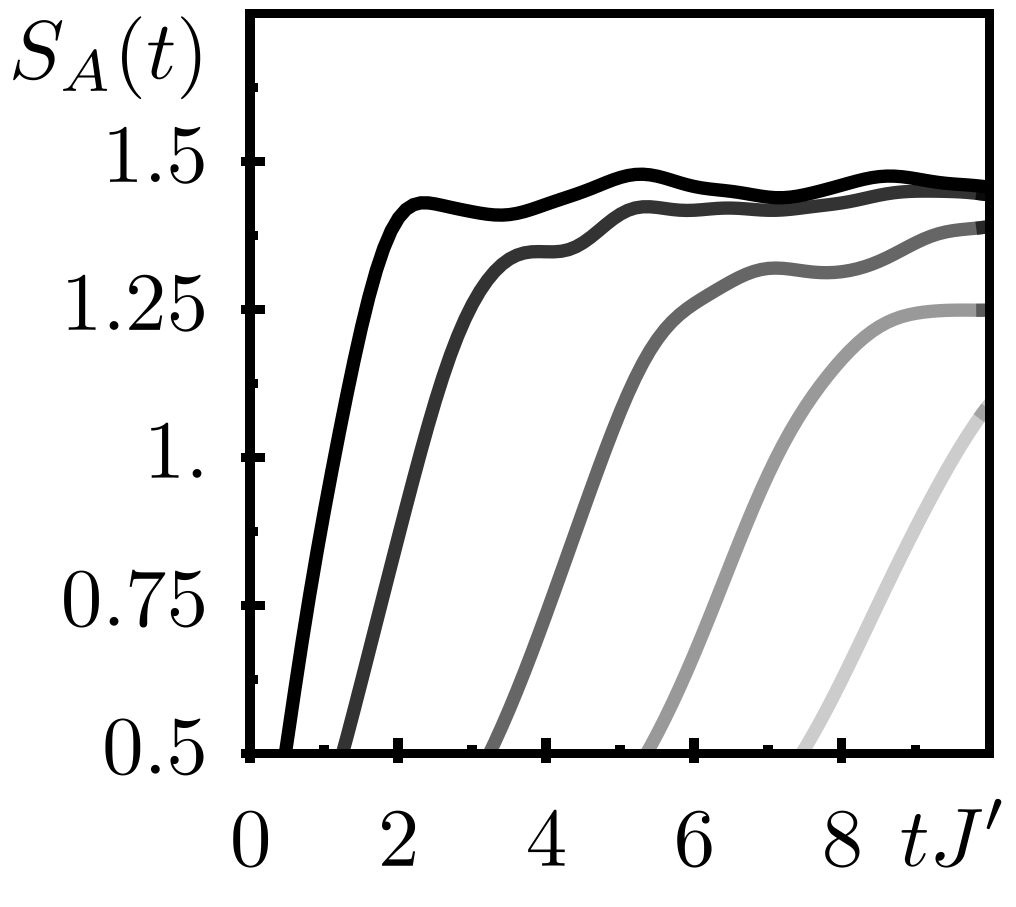}\llap{\parbox[b]{5.8cm}{(d)\\\rule{0ex}{3.3cm}}}\llap{\parbox[b]{5.9cm}{1\\\rule{0ex}{2.3cm}}}\llap{\parbox[b]{5.1cm}{2\\\rule{0ex}{2.3cm}}}\llap{\parbox[b]{3.4cm}{3\\\rule{0ex}{2.3cm}}}\llap{\parbox[b]{1.7cm}{4\\\rule{0ex}{2.3cm}}}\llap{\parbox[b]{1.4cm}{5\\\rule{0ex}{1.3cm}}}
    \caption{Entanglement growth for a local excitation created at $x_0=L/2$ for different bond dimensions and bipartitions. The system consists of $L=32$ unit-cells for a single SSH wire at parameters $J=J'$ without interactions $U_\parallel=U_\perp=0$ (the non-interacting quantum critical point) displayed in panels (a-b) and with $H_{\rm lr}$ interactions $U_\parallel=2$, $U_\perp=0$ shown in (c-d). (a/c) The bipartition is chosen such that $\ell=L/2$ and the excitation is created at the interface between $A|B$. Panels (b/d) demonstrate that increasing the distance between the boundary and the initial location of the excitation (as indicated by the numbers $1,\ldots,5$ within the panels) causes an approximately linear delay of the onset of the rapid growth of the entanglement.
 }
    \label{fig:entanglement_growth}
\end{figure}

In the following, we will estimate the entanglement growth in the underlying system.
It is well-known that a critical system can be represented as a conformal field theory, and methods familiar from renormalization group theory provide an estimate for the asymptotic behavior of such a generic system.
For instance, a global quench of a given (critical) Hamiltonian results in a continuum of quasiparticle excitations spreading through the system (with maximum velocity $v$), and the entanglement $S_{A}(t)$ between a region $A(\ell)$ of the chain of length $\ell$ and the rest of the chain $B(L-\ell)$ will scale as $S_{A}(t) \propto t$ (for times smaller than the characteristic length $t<t_\ell\approx\ell/2v$)~\cite{Calabrese2005}.
Local perturbations, instead, produce a bipartite entanglement which grows only as $S_A(t)\propto \log(t)$ for critical systems, which implies a saturation of $S_A(t)\approx\text{const.}$ if the underlying system is gapped \cite{Calabrese2007}.
Moreover, if we denote by $i$ the distance between the location of the excitation and the $A|B$ bipartition boundary, we expect that the entanglement entropy will suddenly increase {\it after} a critical time proportional to $i$.
In our experiment, we thus expect a rapid saturation of the entanglement entropy in time, making it a perfect candidate to target with tensor network schemes.
We numerically confirm the above statements by the MPS simulations presented in Fig.~\ref{fig:entanglement_growth}.

\begin{figure}[ht]
    \includegraphics{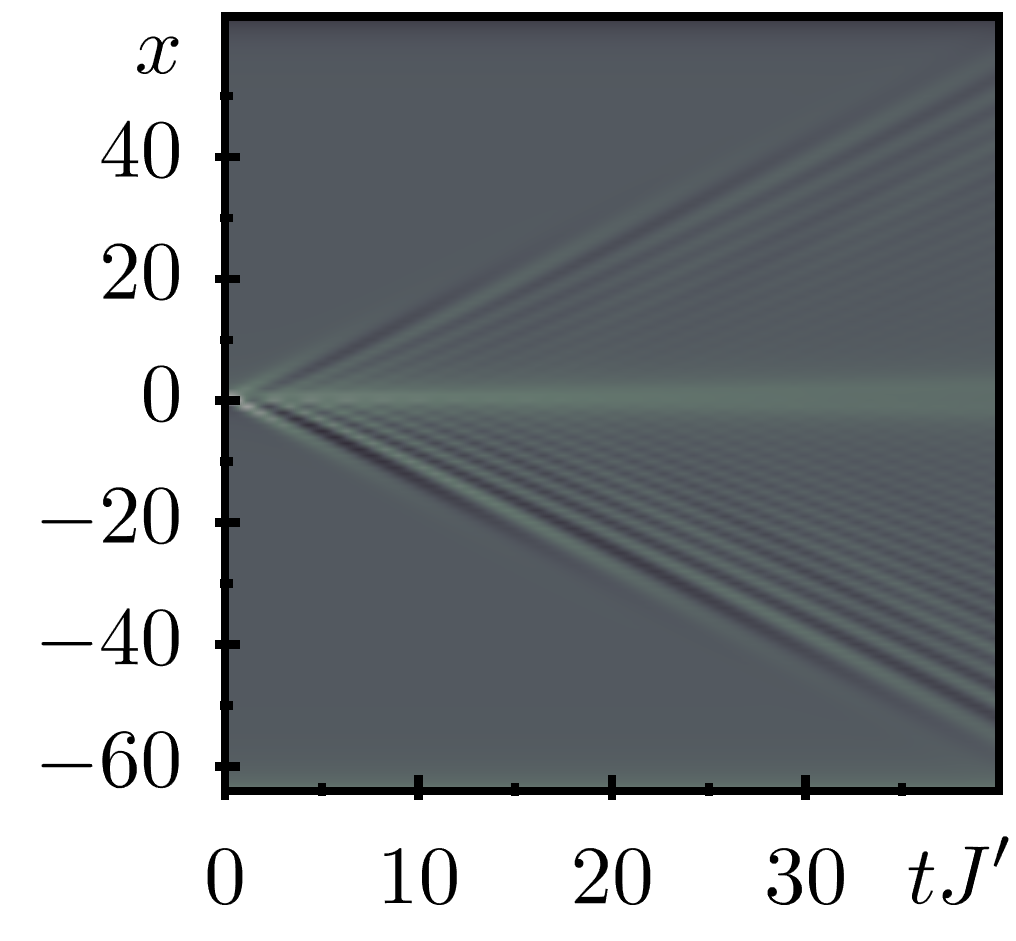}\llap{\parbox[b]{5.8cm}{\color{white}(a)\\\rule{0ex}{3.3cm}}}
    \includegraphics{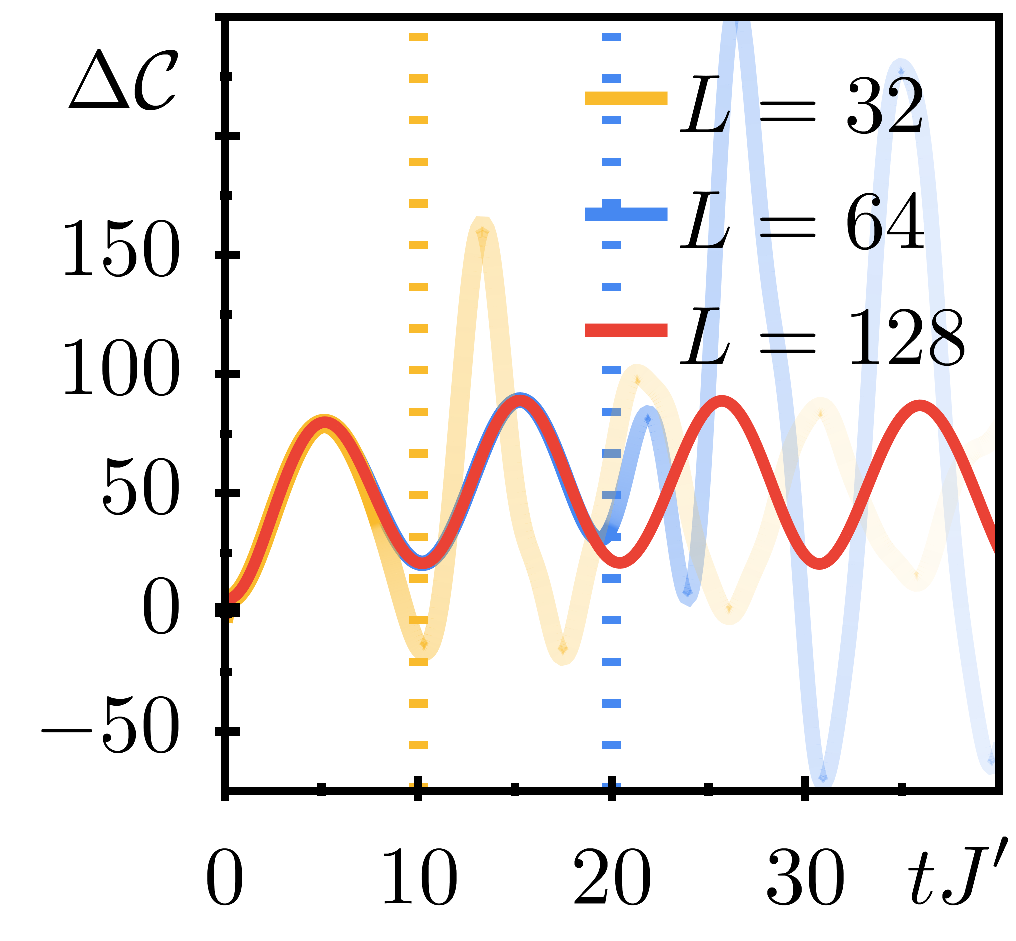}\llap{\parbox[b]{5.8cm}{(b)\\\rule{0ex}{3.3cm}}}
    \includegraphics{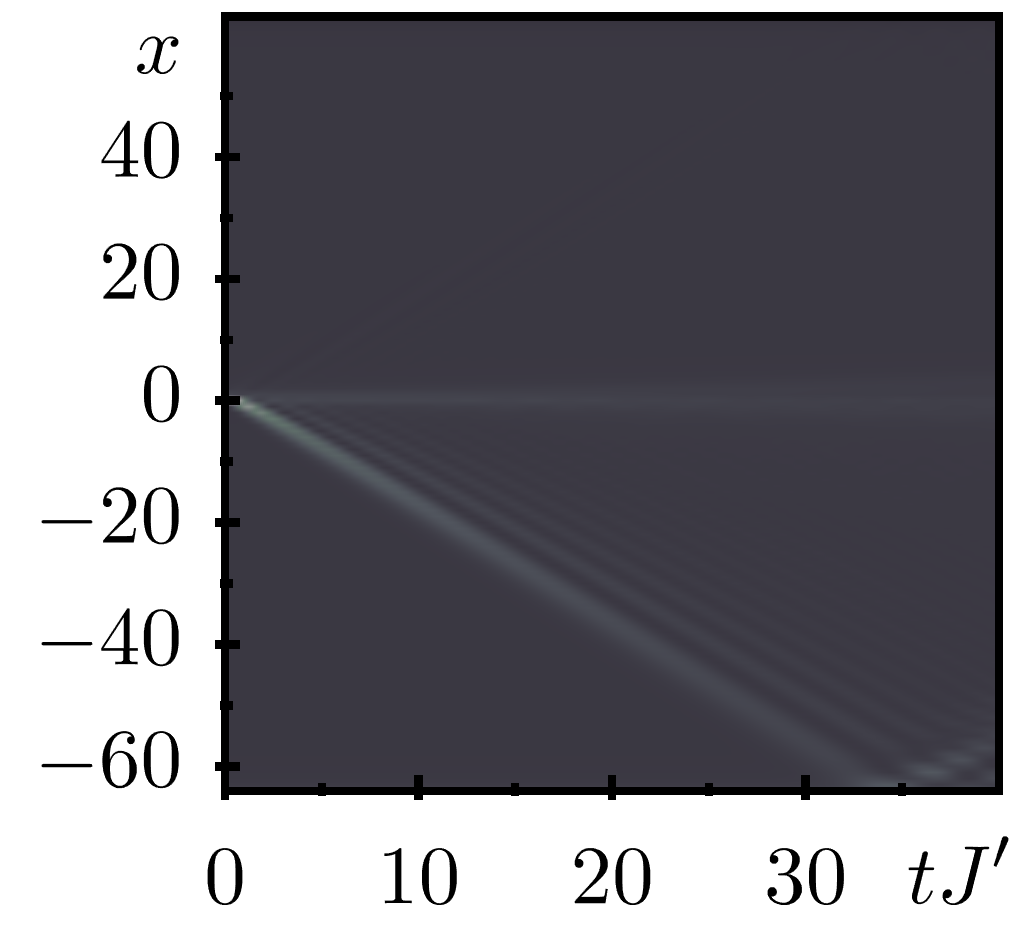}\llap{\parbox[b]{5.8cm}{\color{white}(c)\\\rule{0ex}{3.3cm}}}
    \includegraphics{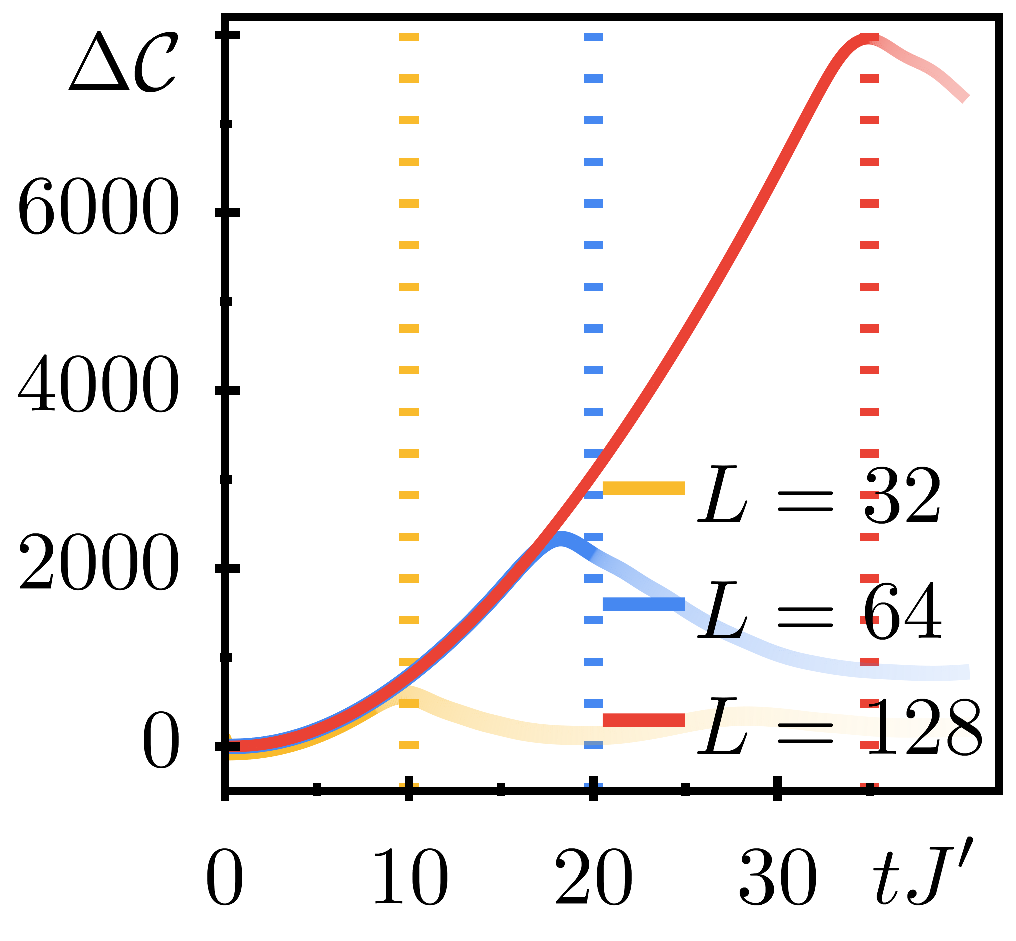}\llap{\parbox[b]{5.8cm}{(d)\\\rule{0ex}{3.3cm}}}
    \caption{
    Panel (a) displays the cone-like spreading of the density of a propagating excitation in the topological (TOI) phase of $\MH_{\rm lr}$ at $J/J'=0.0469$ and $U_\parallel/J'=2.75$ (the yellow triangle in Fig.~\ref{fig:mcd_numerics_lr}(b-c)), computed with bond dimension $M=256$ on a chain of length $L=128$.
    In panel (b) we display corresponding MCD, computed for chains of lengths $L\in\{32,64,128\}$ (yellow, blue, red). The vertical dotted lines highlight the maximum observation time of bulk-propagation for each size.
    The graphs shown in (c-d) illustrate instead the situation in the symmetry-broken (SB) phase $\MH_{\rm lr}$ at $J/J'=0.0469$ and $U_\parallel/J'=5.5$ (gray downward triangle in Fig.~\ref{fig:mcd_numerics_lr}(b-c)). The injection of the excitation induces a spontaneous symmetry breaking, which leads to the formation of the domain-wall visible in panel (c). As a consequence, the MCD displays the characteristic quadratic divergence in time shown in panel (d).}
    \label{fig:MPS_long_time}
\end{figure}

Away from the quantum phase transition lines, the model $H_{\rm lr}$ is gapped such that MPS Ans\"atze are highly efficient: the figure of merit -- the maximum truncated probability in the reduced density matrix of a bipartition -- can be kept at machine precision for finite systems.
This allows us to aim for much larger evolution times and system sizes at individual points in parameter space, which we present in Fig.~\ref{fig:MPS_long_time}.
In particular,  the MCD presents a very irregular behavior close to the phase boundaries, as was visible in Fig.~\ref{fig:mcd_numerics_lr} of the main text.
Here however we show that the MCD becomes more and more regular when one considers larger and larger systems sizes and evolution times.

\begin{figure}[ht]
    \includegraphics{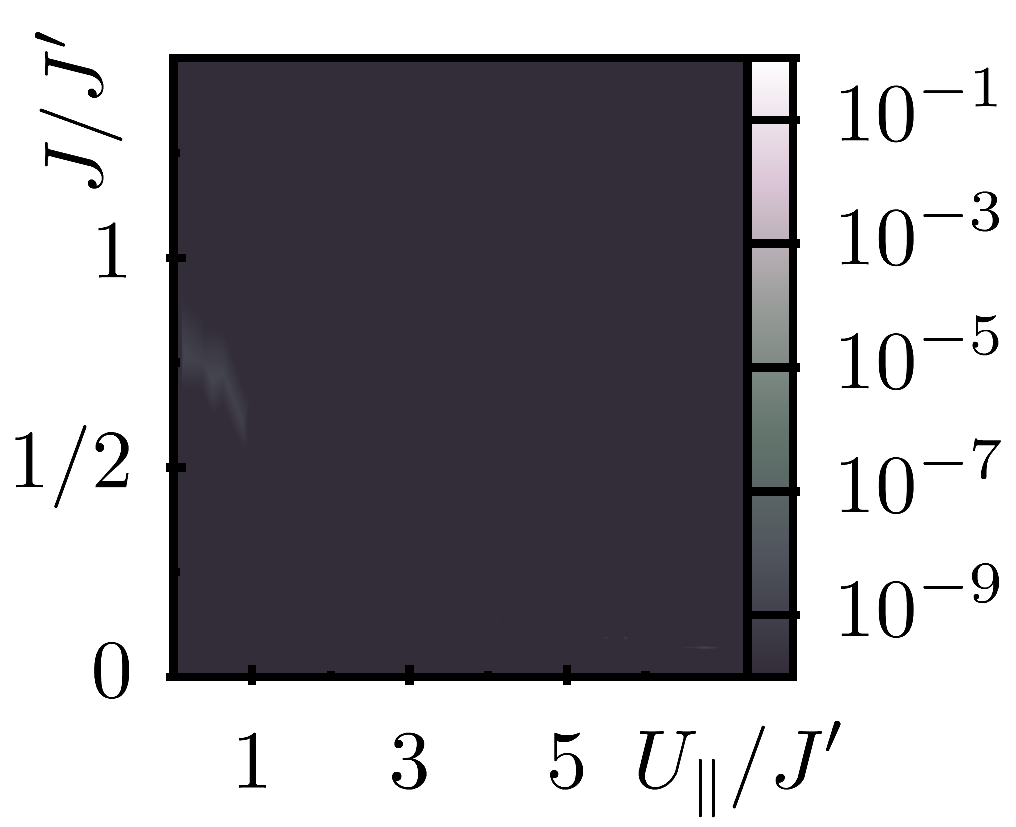}\llap{\parbox[b]{6.5cm}{\color{white}(a)\\\rule{0ex}{2.8cm}}}
    \hfill
    \includegraphics{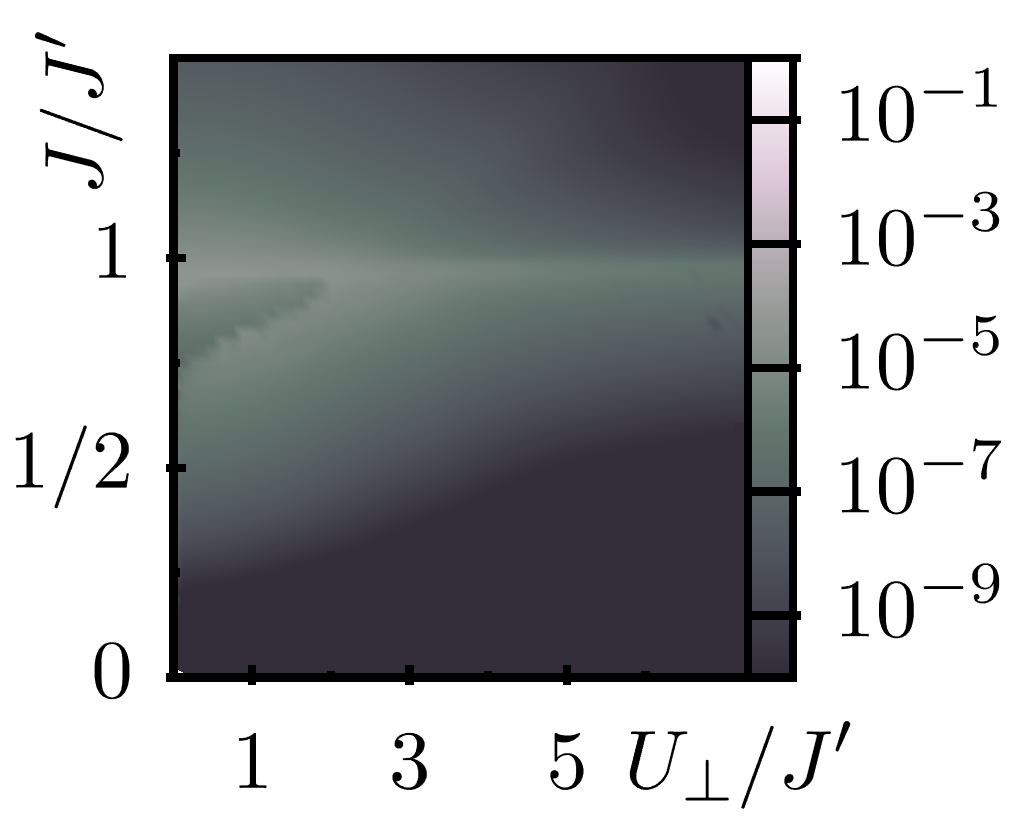}\llap{\parbox[b]{6.5cm}{\color{white}(b)\\\rule{0ex}{2.8cm}}}
    \includegraphics{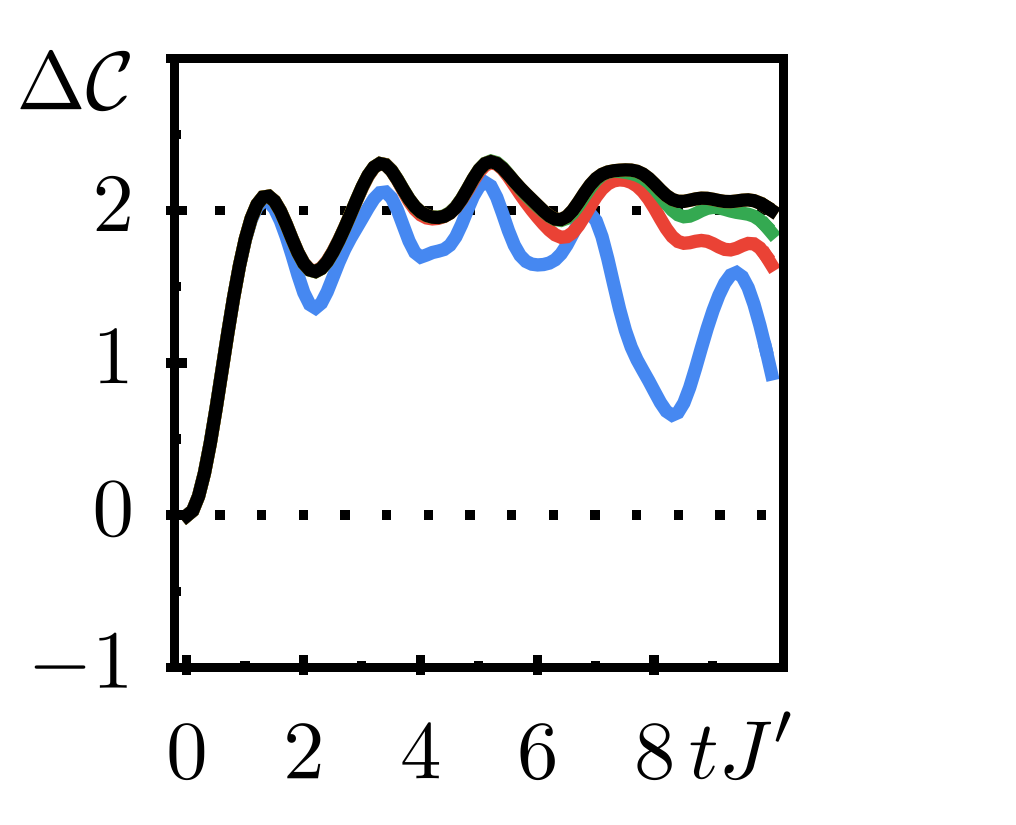}\llap{\parbox[b]{6.5cm}{(c)\\\rule{0ex}{2.8cm}}}
    \hfill
    \includegraphics{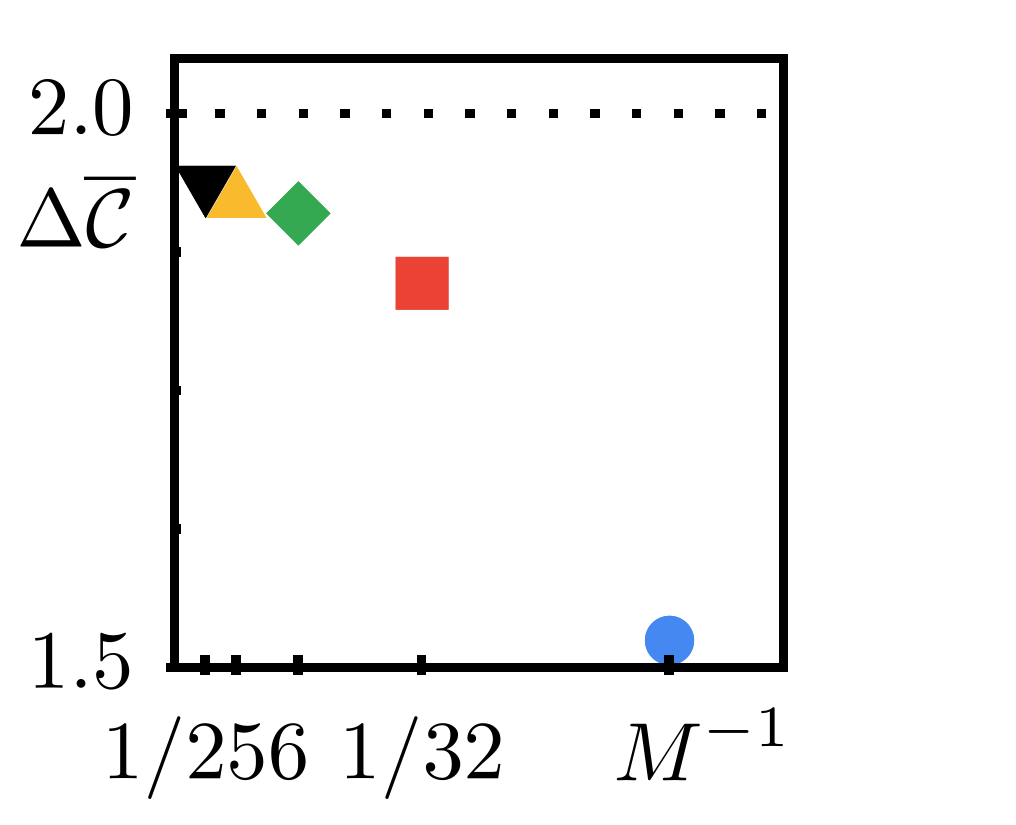}\llap{\parbox[b]{6.5cm}{(d)\\\rule{0ex}{2.8cm}}}
    \includegraphics{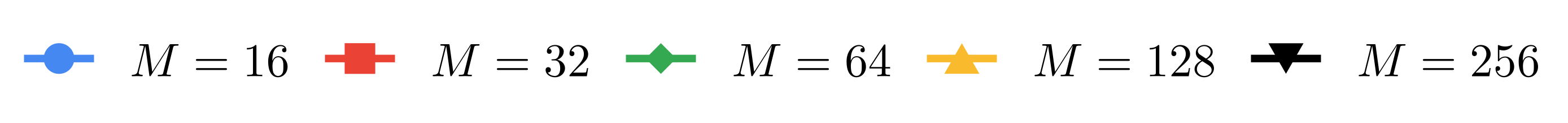}
    \caption{(a) Truncation error in the reduced density matrix of the static ground state approximation for $\MH_{\rm lr}$ for bond dimension $M=64$. The maximum error is $\Delta\rho=1.6\times10^{-9}$, showing that the MPS calculation is essentially exact for $\MH_{\rm lr}$. (b) The same analisys performed on $\MH_{\rm sr}$ shows larger truncation errors, prompting for a check of results vs.~bond-dimension.
    (c) Time-traces of the MCD for $\MH_{\rm sr}$ with $J/J'=0.6$ and $U_\perp=6$ for different bond dimensions. Relatively small bond dimensions ensure to converge within the time-window we investigated ($0<t<8/J'$). (d) The time-averaged $\DCav$ converges therefore very quickly.
    }
    \label{fig:mps_error}
\end{figure}

Although the phase diagram of $\MH_{\rm sr}$ is simpler than $\MH_{\rm lr}$, the two-wire setup is computationally more expensive -- e.g., the ground state at half filling in the topological phase is four-fold degenerate, and the complexity of the gapless region is increased as well.
We compare the numerical complexity to approximate the ground state in Fig.~\ref{fig:mps_error}(a-b).

Combining the precise approximation of the ground state with the accurate time-evolution of a local excitation as confirmed by Fig.~\ref{fig:entanglement_growth}, we conclude that an MPS approach yields negligible numerical inaccuracies for $\MH_{\rm lr}$, whereas errors need to be discussed further in case of $\MH_{\rm sr}$ by means of a bond dimension scaling analysis.
We present the corresponding MCD in Fig.~\ref{fig:mps_error}(c): time-traces visibly differ at the tails of the graph.
Nonetheless,  we show in Fig.~\ref{fig:mps_error}(d) that the time-average of the MCD scales very quickly to the desired precision.
In conclusion, $\DCav$ is quite resistant against numerical errors.
All in all, this provides good confidence that the scheme and results presented in this work are resistant against usual unavoidable inaccuracies of the TN simulations~\cite{Paeckel2019}.
\end{document}